\documentclass[journal]{IEEEtran}

\usepackage[utf8]{inputenc}
\usepackage[T1]{fontenc}
\usepackage{url}
\usepackage{ifthen}
\usepackage{cite}
\usepackage[cmex10]{amsmath}
\usepackage{graphicx}
\usepackage{amsfonts}
\usepackage{amsmath}
\usepackage{amssymb}
\usepackage{mathrsfs}
\usepackage{color}
\usepackage{bm}
\usepackage{latexsym}

\newcommand{\fat}[1]{\mbox{\boldmath$#1$}}%
\newtheorem{theorem}{Theorem}
\newtheorem{corollary}{Corollary}%
\newtheorem{lemma}{Lemma}%
\newtheorem{definition}{Definition}%

\newtheorem{example}{Example}

\newcommand{\myQED}{\mbox{}\hfill{$\Box$}}
\newcommand{\qed}{\hfill $\blacksquare$}

\ifCLASSINFOpdf \else \fi \onecolumn

\begin{document}

\title{Maximum Achievable Rate of Resistive Random-Access Memory
Channels by Mutual Information Spectrum Analysis}

\author{Guanghui Song, \IEEEmembership{Member, IEEE},
        Kui Cai, \IEEEmembership{Senior Member, IEEE},
        Ying Li, \IEEEmembership{Member, IEEE},\\
        and Kees A. Schouhamer Immink, \IEEEmembership{Fellow, IEEE}

\thanks{
    The work of Guanghui Song and Ying Li is supported by the National
    Natural Science Foundation of China (NSFC) under Grants 62131016
    and 62271369.
    The work of Kui Cai is supported by RIE2020 Advanced Manufacturing
    and Engineering (AME) programmatic grant A18A6b0057.
    This work is also supported in part by the National Natural
    Science Foundation of China (NSFC) under Grants 62201424 and 61971333.
     \emph{(Corresponding author: Kui Cai.)}

    Guanghui Song and Ying Li are with  the State Key Lab of
    Integrated Services Networks, Xidian University, Xi'an, 710071,
    China (songguanghui@xidian.edu.cn, yli@mail.xidian.edu.cn).

    Kui Cai is with Science, Mathematics and Technology Cluster,
    Singapore University of Technology and Design, Singapore,
    487372 (cai\_kui@sutd.edu.sg).

    Kees A. Schouhamer Immink is with Turing Machines Inc,
    Willemskade 15d, 3016 DK Rotterdam, The Netherlands
    (immink@turing-machines.com).
} }

\maketitle

\begin{abstract}
The maximum achievable rate is derived for resistive random-access
memory (ReRAM) channel with sneak-path interference. Based on the
mutual information spectrum analysis, the maximum achievable rate
of ReRAM channel with independent and identically distributed
(i.i.d.) binary inputs is derived as an explicit function of
channel parameters such as the distribution of cell selector
failures and channel noise level.  Due to the randomness of cell
selector failures, the ReRAM channel demonstrates multi-status
characteristic. For each status, it is shown that as the array
size is large, the fraction of  cells  affected by sneak paths
approaches a constant value. Therefore, the mutual information
spectrum of the ReRAM channel is formulated as a mixture of
multiple stationary channels. Maximum achievable rates  of the
ReRAM channel with different settings, such as  single- and
across-array codings, with and without data shaping, and optimal
and treating-interference-as-noise (TIN) decodings, are compared.
These results provide valuable insights on the code design for
ReRAM.
\end{abstract}


\section{Introduction}
\IEEEPARstart{A}{s} an emerging non-volatile memory technology,
resistive random-access memory (ReRAM)  has demonstrated many
advantages as compared to conventional memory, such as ultra-dense
data storage and parallel reading and writing
\cite{Strukov,Chen,Zahoor,Wong}. However,  its cross-bar structure
causes a severe interference problem called ``sneak path"
\cite{Zidan,Naous,Luo1,Kvatinsky,Yaakobi,Yuval}, which leads to a
high inter-cell signal correlation. Recent works showed that the
sneak path occurrence is highly related to the selector failures
(SFs) in the resistive memory arrays
\cite{Ben,CZH,SongArxiv,SongTcom}, and efficient data detection
and coding strategies are proposed based on random SF models.
Ben-Hur and Cassuto \cite{Ben} proposed a single-cell data
detection scheme without considering the inter-cell correlation.
Chen \emph{et al.} developed a pilot-bit-assisted data detection
scheme, which can utilize part of the inter-cell correlation. Our
previous work \cite{SongArxiv} proposed a near-optimal joint data
and sneak path detection scheme which can fully take advantage of
the inter-cell correlation. To further reduce the detection error
rate, we proposed error correction codes for ReRAM
\cite{SongTcom,Panpan1,Panpan2,SunTcom}. An across-array coding
strategy with soft message-passing decoding and irregular
repeat-accumulate (IRA) code design was proposed in
\cite{SongTcom}. Works \cite{Panpan1,Panpan2} proposed low density
parity check (LDPC) codes for ReRAM channel with and without
quantization. Work \cite{SunTcom} applied polar code to ReRAM and
proposed a belief propagation based joint detection and decoding
to deal with both the sneak-path interference and channel noise.

So far, there are very limited works on achievable rate analysis
of the ReRAM channel.  Cassuto  \emph{et al.} \cite{Yuval}
analyzed the achievable information rate of the noiseless ReRAM
channel with constrained data input (to avoid sneak path). It is
shown that even for a noiseless channel, to completely avoid sneak
path, the maximum achievable rate tends to 0 as the memory array
size tends to infinity. Our previous work  \cite{SongTcom}
provided an upper and lower bounds of the capacity of noisy ReRAM
channel under a treating-interference-as-noise (TIN) assumption,
where the sneak-path interference is considered as independent and
identically distributed (i.i.d.) noise during decoding. Thus, the
channel in each memory array is regarded as a memoryless channel.
However, the exact achievable rate of the ReRAM channel under
optimal decoding is still unknown.

The main challenge is that the sneak-path interference within each
memory array is correlated and hence the ReRAM channel is a
channel with memory. Existing information theory that deals with
such channels is the mutual information spectrum method with the
general formula of channel capacity \cite{Verdu,HanBook}. However,
the analysis of the information spectrum of a channel is usually
not easy.

In this paper, we propose a maximum achievable rate analysis for
the ReRAM channel using the mutual information spectrum method.
The maximum achievable rate of a ReRAM channel with i.i.d. binary
inputs is derived as an explicit function of channel parameters
such as the  channel noise level and the distribution of cell SFs.
The randomness of cell SFs results in multi-status characteristics
of the ReRAM channel. For instance, more SFs produce more sneak
paths in an array resulting in a worse channel, while fewer SFs
lead to a better channel. We prove that as the array size becomes
large, the fraction of cells affected by sneak paths approaches to
a constant value that depends on the number of SFs in the array
and the channel input distribution. The mutual information
spectrum of the ReRAM channel is finally formulated as a mixture
of that of multiple stationary channels. Moreover, maximum
achievable rates of the ReRAM channel are compared under different
settings, such as single- and across-array codings, with and
without data shaping, and optimal and TIN decodings. These results
provide the following insights on the code design for ReRAM.

\begin{itemize}
\item \emph{Single-array coding:} If the data of each memory array
is independently encoded, the maximum achievable rate is subject
to the worst case of channel status, which is determined by the
maximum possible number of SFs in a memory array.

\item \emph{Across-array coding:} By jointly encoding the data of
multiple memory arrays, the ergodic information rate which does
not subject to the worse channel status is achievable. Therefore,
compared with the single-array coding, a rate gain can be obtained
by across-array coding. This gain is called a diversity gain
because it essentially originates from multiple independent
channel observations of a codeword.

\item  \emph{Data shaping:}  The maximum information rate is
achieved by an asymmetric input data distribution. A remarkable
shaping gain can be achieved by optimizing the input data
distribution. This is different from the traditional memoryless
channels. It is known that for memoryless channels, the symmetric
input usually achieves the maximum or near maximum rates.

\item  \emph{Joint data and sneak path detection:} The optimal
decoder that can take advantage of the channel correlation
achieves much higher rate than the sub-optimal TIN decoder.
Therefore, joint data and sneak path detection is desirable if the
corresponding complexity is affordable to the system.
\end{itemize}

The above insights coincide with simulation results observed in
\cite{SongArxiv} and \cite{SongTcom}. Therefore, this work
provides a theoretical support for the results of \cite{SongArxiv}
and \cite{SongTcom}.

The rest of this paper is organized as follows.
Section~\ref{sec:model} presents the sneak path model of ReRAM. In
Section~\ref{sec:main results}, we define the ReRAM channel and
give the main results of this paper. Section~\ref{sec:proof}
proposes the mutual information spectrum analysis which serves as
the proof for Theorem~\ref{thm:main1}. The paper is concluded by
Section~\ref{sec:conclude}.

\begin{figure*}[t]
    \includegraphics[width=
    5.6 in]{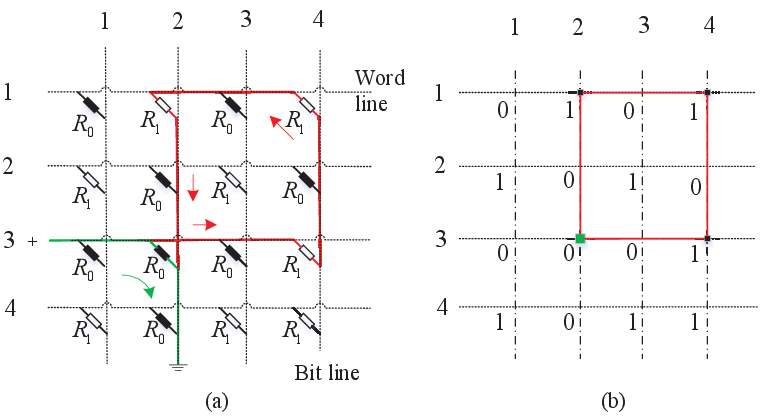}
    \centering
    \caption{Sneak path during the reading of cell $(3, 2)$ in a
    $4\times 4$ memory array. (a) is the memory array and (b) is
    the corresponding data array. The green line is the desired
    current path for resistance measuring and the red line going
    through cells $(3, 2)\rightarrow(3, 4)\rightarrow(1, 4)
    \rightarrow(1, 2)\rightarrow(3, 2)$ is a sneak path.
    Note that word lines and bit lines are connected via memory
    cells. Arrows show current flow directions. A reverse current
    flows across cell $(1, 4)$.} \label{fig:SPmodel}
\end{figure*}

\section{Sneak Path Model}\label{sec:model}

Consider the storage of an $N\times N$ binary data array
$x^{N\times N}=[x_{m,n}]_{N\times N}, x_{m,n}\in\{0, 1\}$, by a
ReRAM array with $N^2$ memory cells. A memory cell is a memristor
which has two resistance states, i.e., the High-Resistance State
(HRS) with resistance value $R_0$ and the Low-Resistance State
(LRS) with resistance value $R_1$, corresponding to the two states
logical-0 and -1 of a bit. The memory cells are arranged in a
cross-bar structure, where cell $(m, n)$ that lies at the
intersection of the $m$-th row (word line) and the $n$-th column
(bit line) is used to store a bit of $x_{m,n}$ for $m=1,...,N,
n=1,...,N$. During memory reading, the resistance value of each
memory cell is detected to determine the stored data bit. However,
due to the existence of  sneak paths and various channel noises,
this process is quite unreliable
\cite{Ben,CZH,SongTcom,SongArxiv}.

When cell $(m, n)$ is read, the $m$-th word line is biased with a
voltage and the $n$-th bit line is grounded so that current goes
through cell $(m, n)$ and its resistance is detected. A sneak path
is an undesired current path in parallel to the desired current
path for resistance measuring. It is defined as a closed path that
originates from and returns to cell $(m, n)$ while traversing
logical-1 cells through alternating vertical and horizontal steps.
Fig.~\ref{fig:SPmodel} shows an example of sneak path during the
reading of cell $(3, 2)$. The green line shows the desired path
for resistance measuring and the red line going through cells $(3,
2)\rightarrow(3, 4)\rightarrow(1, 4)\rightarrow(1,
2)\rightarrow(3, 2)$ is a sneak path. A direct impact of this
sneak path is that it lowers the detected resistance value. In
this case, the detected resistance value is
\begin{equation}
R_0^\prime=\left(\frac{1}{R_0}+\frac{1}{R_s}\right)^{-1}<R_0,
\end{equation}
where $R_s$ is a parasitic resistance value caused by the sneak
path. Since the sneak path lowers the detected resistance value,
it is only harmful when a cell with logical-0 is read and makes it
more vulnerable to noise.  Hence we ignore its influence to the
logical-1 cells.

The most popular method to mitigate sneak paths  is to introduce a
selector in series to each array cell. A cell selector is an
electrical device that allows current to flow only in one
direction across the cell. Since sneak paths inherently produce
reverse current in at least one of the cells along the parallel
path (e.g. cell $(1, 4)$ in Fig.~\ref{fig:SPmodel}), cell
selectors can completely eliminate sneak paths from the entire
array.  However, one basic assumption is that imperfections in the
production or maintenance of the memory cause cell selectors to
fail, leading to reoccurrence of sneak paths
\cite{Ben,CZH,SongArxiv,SongTcom}. In the case of
Fig.~\ref{fig:SPmodel}, due to the circuit structure of the
crossbar array, cells $(3, 4)$ and $(1, 2)$ will conduct current
in the forward direction and not be affected by their selectors.
Only when the selector of cell $(1, 4)$ is faulty will a sneak
path be formed.

If a sneak path occurs during the reading of cell $(m,n)$, we call
cell $(m,n)$ an SP cell.  Following previous works
\cite{Ben,CZH,SongTcom,SongArxiv}, we consider the one diode-one
resistor (1D1R) type selectors and model this behavior using a
random fault model. According to
\cite{Ben,CZH,SongTcom,SongArxiv},  cell $(m,n)$ is an SP cell if
and only if the following three conditions are satisfied:

1) $x_{m, n}=0$.

2) We can find at least one combination of $i, j\in\{1,...,N\}$
that satisfies
\begin{equation}
x_{m, j}=x_{i, j}=x_{i, n}=1.\nonumber
\end{equation}

3) The selector at the diagonal cell $(i, j)$ fails.

The above sneak path occurrence conditions limit the sneak paths
to a length of 3, i.e., traversing three cells. Following
\cite{Ben,CZH,SongTcom,SongArxiv}, we ignore longer sneak paths
since compared to the length-3 sneak paths their affection is
insignificant. Also following \cite{CZH,SongTcom,SongArxiv}, we do
not consider the superposition effect of multiple sneak paths. A
more sophisticated sneak path model was considered in \cite{Ben},
and the principle of our work can be extended to the model in
\cite{Ben}.

We define a sneak path indicator $v_{m, n}$ for cell $(m, n)$ to
be a Boolean variable with $v_{m, n}=1$ if cell $(m, n)$ is an SP
cell, otherwise, $v_{m, n}=0$. Building on the above  sneak path
occurrence conditions, we write the readback signal array
$y^{N\times N}=[y_{m,n}]_{N\times N}$ obtained by the memory
reading as
\begin{equation}
y_{m,n}=\begin{cases} (\frac{1}{R_0}+\frac{v_{m,n}}{R_s})^{-1}
+z_{m,n}&\mbox{ $x_{m,n}=0$},\\  R_1+z_{m,n}&\ \mbox{$x_{m,n}=1$},
\end{cases}
\label{eq:readback}
\end{equation}
where $z_{m,n}, m=1,...,N, n=1,...,N$ are samples of i.i.d.
Gaussian noise with mean 0 and variance $\sigma^2$. This provides
a basic readback signal model of ReRAM.  Variations of
\eqref{eq:readback} for ReRAM with lognormal distributed
resistance variations and channel quantization can be found in
\cite{Panpan1,Panpan2}.

For a given memory array, let
\begin{equation}
\varphi=\left\{(i, j)|\  \textrm{selector at cell}\ (i, j)\
\textrm{fails}\right\}
\end{equation}
be the set that includes all the indices of the locations of
failed selectors. We call $\varphi$ the SF pattern of the array.
Note that $\varphi$ is unknown to the data detector.

\begin{definition} \label{def:scattered}
$\varphi$ is a scattered SF pattern if it only contains the
indices of failed selectors that lie in different rows and
columns, i.e., for any $(i_1,j_1), (i_2,j_2)\in\varphi$ with
$(i_1,j_1)\neq(i_2,j_2)$, we have $i_1\neq i_2$ and $j_1\neq j_2$.
\end{definition}

\begin{definition}
For $(i, j)\in\varphi$, row $i$ and column $j$ of $\fat{X}$ are
called  SF row and column, respectively. A row or column that is
not an SF row or column is a non-SF row or column.
\end{definition}

\begin{figure}[t]
\includegraphics[width=
3.2 in]{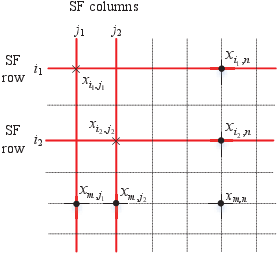} \centering \caption{A memory array with SF
pattern $\varphi=\{(i_1,j_1), (i_2, j_2)\}$. The readback signal
$R_{x_{m,n}}\left((x_{m,j_1}x_{i_1,j_1}x_{i_1,n})\bigcup
(x_{m,j_2}x_{i_2,j_2}x_{i_2,n})\right)$ of cell $(m,n)$ (without
noise) is determined by data stored at the cell $x_{m,n}$, data at
the SF locations $x_{i_1,j_1}, x_{i_2,j_2}$, and data in the SF
rows and columns $x_{i_1,n}, x_{i_2,n}$ and $x_{m,j_1},
x_{m,j_2}$.} \label{fig:readout-signal}
\end{figure}

Define
\begin{equation}
R_x(v)=\left(\frac{1}{R_x}+(1-x)\frac{v}{R_s}\right)^{-1},\ \
\textrm{for} \ x, v\in\{0, 1\}.
\end{equation}
We have $R_1(0)=R_1(1)=R_1$ and $R_0(0)=R_0, R_0(1)=R_0^\prime$.
Based on the sneak path occurrence condition, the readback signal
in (\ref{eq:readback}) can be rewritten as
\begin{eqnarray}
y_{m,n}\!\!\!\!\!\!\!\!\!\!&&= R_{x_{m,n}}\left(v_{m,n}\right)+z_{m,n}\\
&&=
R_{x_{m,n}}\left(\bigcup_{(i,j)\in\varphi}x_{m,j}x_{i,j}x_{i,n}\right)
+z_{m,n},\label{eq:readback2}
\end{eqnarray}
where $\bigcup$ is the logical OR operator, i.e.,
$\bigcup_{(i,j)\in\varphi}x_{m,j}x_{i,j}x_{i,n}=1$ if at least one
of $(i,j)\in\varphi$ with $x_{m,j}x_{i,j}x_{i,n}=1$, otherwise,
$\bigcup_{(i,j)\in\varphi}x_{m,j}x_{i,j}x_{i,n}=0$.
Eq.~(\ref{eq:readback2}) indicates that the readback signal of
cell $(m,n)$ (without noise) is a function of $x_{m, n}$, SF
pattern $\varphi$, and the entries of the SF rows and columns. For
example, in the memory array shown in
Fig.~\ref{fig:readout-signal}, the readback signal of cell $(m,n)$
is related to 7 entries in the data array causing high inter-cell
correlations. Since the SF located at $(i,j), (i,j)\in\varphi$,
may cause sneak paths only when the cell has LRS $x_{i,j}=1$, in
this case, the SF at $(i,j)$ is referred to as an active SF.

\section{ReRAM Channel and Main Results}\label{sec:main results}
In this section, we define a ReRAM channel and provide the main
results of this paper. We first define some notations that will be
used in the rest of this paper. Random variables are denoted by
upper case letters, such as $X$, and their realizations (sample
values) by lower case letters, such as $x$. Random data arrays
with dimension  $N\times N$ are denoted in the form of $X^{N\times
N}$ and $X_{m,n}$ is the $(m,n)$-th entry. Similarly, the
corresponding sample values are denoted as $x^{N\times N}$ and
$x_{m,n}$. For a random variable $X$, its probability density
function is denoted as $P_X(x)$. The  joint  and conditional
probability functions of $X,Y$ are denoted as $P_{X,Y}(x,y)$ and
$P_{Y|X}(y|x)$. We have similar notations for random arrays, such
as $P_{X^{N\times N}}(x^{N\times N})$, $P_{X^{N\times
N},Y^{N\times N}}(x^{N\times N},y^{N\times N})$  and
$P_{Y^{N\times N}|X^{N\times N}}(y^{N\times N}|x^{N\times N})$.
Let $\mathbb{E}$ be an event, its complementary event is denoted
by $\mathbb{E}^c$. For set $A$, $|A|$ is its cardinality. We use
$\lim_{N\rightarrow\infty}X_N \overset{p}{=} X$ to denote that
random process $X_N, N=1, 2, ..., \infty$, converges in
probability to random variable $X$.

Eq.~(\ref{eq:readback2}) specifies a ReRAM channel with input
$x^{N\times N}$ and output $y^{N\times N}$. The channel status of
each memory array is strongly related to the cardinality of its SF
pattern $\varphi$. If $\varphi$ is empty, $|\varphi|=0$, the
channel becomes an i.i.d. binary input additive Gaussian channel.
If $|\varphi|>0$, the channel is correlated due to the sneak path
interference. The number of SP cells is determined by both of the
cardinality of SF pattern $|\varphi|$ and the input data pattern.
Intuitively, larger $|\varphi|$ leads to higher chance to have SP
cells, and hence a worse channel.

Since selectors fail with very low probabilities
\cite{Ben,CZH,SongTcom,SongArxiv}, the number of SFs in an array
is usually very small. To reflect this fact, we assume the number
of SFs in each array  is bounded by a constant $K$. Define
$\Omega_N^k$ as a $k$-SF pattern set, which includes all the SF
patterns of cardinality $k$. Hence $\Omega_N^k$ includes all the
subsets of $\{(m,n)|m=1,...,N,n=1,..,N\}$ with cardinality $k$.
Further define $\Omega_N=\cup_{k=0}^K\Omega_N^k$ as the universal
set of all the possible SF patterns in an array. From the
definitions, we have $|\Omega_N^k|=\binom{N^2}{k}$ and
$|\Omega_N|=\sum_{k=0}^K\binom{N^2}{k}$. We define a probability
distribution $\mathcal{P}=\{p_1, p_2,...,p_K\}$ with $p_K>0,
\sum_{k=1}^Kp_k=1$, upon the SF pattern space $\Omega_N$, where
$p_k$ is the total probability of SF patterns belonging to
$\Omega_N^k$. By assuming equiprobability of SF patterns in
$\Omega_N^k$, each SF pattern in $\Omega_N^k$ occurs with
probability $p_k/\binom{N^2}{k}$. Let $\Phi$ be a random variable
of SF pattern $\varphi$, and $X^{N\times N}$, $Y^{N\times N}$, and
$Z^{N\times N}$ are random variable arrays with sample values of
$x^{N\times N}$, $y^{N\times N}$, and $z^{N\times N}$. We have the
following definition.

\begin{definition} \label{def:ReRAM}
ReRAM channel $W^\Phi_N(\cdot|\cdot): \{0, 1\}^{N\times
N}\rightarrow \mathbb{R}^{N\times N}$ is a channel with input
$X^{N\times N}$ and output $Y^{N\times N}$  that satisfies
\begin{equation}
Y_{m,n}=R_{X_{m,n}}\left(\bigcup_{(i,j)\in\Phi}X_{m,j}X_{i,j}X_{i,n}\right)
+Z_{m,n},\label{eq:Y}
\end{equation}
where $P_\Phi(\varphi)=p_k/\binom{N^2}{k}$ for $\varphi\in
\Omega_N^k, k=0,1,...,K$, and $Z_{m,n}\sim \mathcal{N}(0,
\sigma^2), m=1,...,N, n=1,...,N$ are i.i.d. Gaussian random
variables.
\end{definition}

Definition~\ref{def:ReRAM} captures the random nature of ReRAM
channel caused by the random SF patterns in memory arrays. We
investigate its maximum achievable coding rate with input data
distribution of i.i.d. Bernoulli $(q), 0<q<1$ with
$\textrm{Pr}(X_{m,n}=1)=q$ and $\textrm{Pr}(X_{m,n}=0)=1-q$ for
$i=1,..., N, j=1,..., N$. The main technique we use here is a
mutual information spectrum analysis, which is a tool to analyze
the achievable rate of  non-i.i.d. channels including
non-stationary, non-ergodic, and channels with memory
\cite{HanBook,Verdu}.

\begin{definition}\label{def:I}
Mutual information density rate between input $X^{N\times N}$ and
output $Y^{N\times N}$ of ReRAM channel $W^\Phi_N(\cdot|\cdot)$ is
\begin{equation}
\mathcal{I}_q^N=\frac{1}{N^2}\log\frac{W^\Phi\left(Y^{N\times N}
|X^{N\times N}\right)}{P_{Y^{N\times N}}\left(Y^{N\times
N}\right)}. \label{eq:Iq}
\end{equation}
The probability density function $P_{\mathcal{I}_q^N}$ of
$\mathcal{I}_q^N$ is mutual information spectrum. Note that for
finite $N$, $\mathcal{I}_q^N$ can be a real number, while we can
show that, as $N\rightarrow\infty$, $\mathcal{I}_q^N$ will be in
the range of $(0, 1]$ with probability 1.
\end{definition}

In general, $\mathcal{I}_q^N$ is a random variable related to the
parameter $q$ of the input distribution. We assume a
\textit{single-array coding} strategy where the data of each array
is encoded as a single codeword of length $N^2$, and data of
different arrays are encoded independently as different codewords.
As the code length approaches infinity, the decoding error
probability over the ReRAM channel can be arbitrary small if and
only if $r\leq\underline{\mathcal{I}}_q$, where
\begin{equation}
\underline{\mathcal{I}}_q=\sup\left\{x|\lim_{N\rightarrow\infty}
\textrm{Pr}(\mathcal{I}_q^N<x)=0\right\}\nonumber
\end{equation}
is the spectral  inf-mutual information rate \cite{HanBook,Verdu},
and $r$ is the code rate. By optimizing $q$, we obtain the maximum
achievable rate.
\begin{definition}\label{def:R}
The maximum achievable rate $\mathcal{R}$ of ReRAM channel
$\{W^\Phi_N(\cdot|\cdot)\}_{N=1}^\infty$ with arbitrary small
error probability is
\begin{equation}
\mathcal{R}=\sup_q\underline{\mathcal{I}}_q.
\end{equation}
\end{definition}

Suppose as $N\rightarrow\infty$, $\mathcal{I}_q^N$ converges in
probability to random variable $\mathcal{I}_q^\infty$, which is
the asymptotic mutual information rate. According to
Definition~\ref{def:R}, the maximum achievable rate of the ReRAM
channel is
$\mathcal{R}=\sup_q\sup\left\{x|\textrm{Pr}(\mathcal{I}_q^\infty<x)
=0\right\}$. This is a generalization of the conventional i.i.d.
channel, whose asymptotic mutual information rate is a constant as
shown in the following example.

\begin{example}[$\Phi=\phi$] \label{eg:iid}
Consider the trivial case  of  $\Phi=\phi, N=1, 2,...,\infty$,
where $\phi$ is the empty set. In this case, the ReRAM channel
becomes a conventional i.i.d. binary input additive Gaussian
channel with
\begin{eqnarray}
\mathcal{I}_q^\infty&&\!\!\!\!\!\!\!\!\!\!=
\lim_{N\rightarrow\infty}\mathcal{I}_q^N\nonumber\\
&&\!\!\!\!\!\!\!\!\!\!=\lim_{N\rightarrow\infty}\frac{1}{N^2}
\sum_{m=1}^N\sum_{n=1}^N\log\frac{P_{Y_{m,n}|X_{m,n}}
\left(Y_{m,n}|X_{m,n}\right)}{P_{Y_{m,n}}\left(Y_{m,n}\right)}
\label{eq:proudP}\\
&&\!\!\!\!\!\!\!\!\!\!\overset{p}{=}\textrm{E}
\left[\log\frac{P_{Y_{1,1}|X_{1,1}}\left(Y_{1,1}|X_{1,1}\right)}
{P_{Y_{1,1}}\left(Y_{1,1}\right)}\right]\label{eq:E_XY}\\
&&\!\!\!\!\!\!\!\!\!\!=I(X_{1,1},Y_{1,1}), \label{eq:I_XY}
\end{eqnarray}
where (\ref{eq:proudP}) and (\ref{eq:E_XY}) are due to the i.i.d.
property and the law of large numbers, and $I(X_{1,1},Y_{1,1})$ is
a symbol-wise mutual information. In this case, the mutual
information density rate is a constant. Let $\lambda_1=q,
\lambda_0=1-q$ be the input data distribution. Let
$\gamma=(R_0-R_1)/(2\sigma)$, and
$f(y,\mu,\sigma)=\frac{1}{\sqrt{2\pi}\sigma}e^{-\frac{(y-\mu)^2}{2\sigma^2}}$
be the  Gaussian probability density function of mean $\mu$ and
standard deviation $\sigma$.  Using
$P_{Y_{1,1}}(Y_{1,1})=\sum_{x=0,1}\lambda_xf(Y_{1,1},R_x,\sigma)$
and $P_{Y_{1,1}|X_{1,1}}(Y_{1,1}|X_{1,1})
=f(Y_{1,1},R_{X_{1,1}},\sigma)=f(Z_{1,1},0,\sigma)$,
(\ref{eq:I_XY}) is derived as a function of $q$ and $\gamma$:
\begin{eqnarray}
&&\!\!\!\!\!\!\!\!\!\!I(X_{1,1},Y_{1,1})\nonumber\\
=&&\!\!\!\!\!\!\!\!\!\! \textrm{E}\left[\log\frac{1}{\sum_{x=0,1}
\lambda_xf(Y_{1,1},R_x,\sigma)}\right]-
\textrm{E}\left[\log\frac{1}{f(Z_{1,1},0,\sigma)}\right]\nonumber\\
=&&\!\!\!\!\!\!\!\!\!\!-\int_{-\infty}^{+\infty}\sum_{x=0,1}
\lambda_xf(y,R_x,\sigma)\log\sum_{x=0,1}\lambda_xf(y,R_x,\sigma)dy\nonumber\\
&&\!\!\!\!\!\!\!\!\!\!\ \ \ \ \ \ \ \ -\log\sqrt{2\pi e\sigma^2}
\label{eq:Hz}\\
=&&\!\!\!\!\!\!\!\!\!\!-\int_{-\infty}^{+\infty}\sum_{x=0,1}
\lambda_xf(y,(-1)^x\gamma,1)\log\sum_{x=0,1}\lambda_xf(y,(-1)^x\gamma,1)
dy\nonumber\\
&&\!\!\!\!\!\!\!\!\!\!\ \ \ \ \ \ \ \ -\log\sqrt{2\pi e}\nonumber\\
\triangleq&&\!\!\!\!\!\!\!\!\!\! C_q(\gamma),
\end{eqnarray}
where we used $\textrm{E}\left[Z_{11}^2/\sigma^2\right]=1$ in
(\ref{eq:Hz}). Since $\gamma$ is in fact the square root of
channel signal-to-noise ratio (SNR), given $q$, $C_q(\gamma)$ is a
monotone increasing function of $\gamma$. For a given $\gamma$,
$C_q(\gamma)$ is maximized at $q=0.5$, which is known for the
binary-input additive Gaussian channel \cite{LinBook}. In this
case the mutual information spectrum as $N$ approaches infinity is
a Dirac delta function of
$P_{\mathcal{I}_q^\infty}(x)=\delta[x-C_q(\gamma)]$. The maximum
achievable rate  is hence
$\mathcal{R}=\sup_q\underline{\mathcal{I}}_q
=\sup_qC_q(\gamma)=C_{0.5}(\gamma)$. It can be shown that
$C_{0.5}(\gamma)$ is exactly the channel capacity, the maximum
rate achieved by any input distribution including non-i.i.d. input
distributions.\qed
\end{example}

For the general ReRAM channel described by
Definition~\ref{def:ReRAM} with a random SF pattern, letting
$\gamma^\prime=(R_0^\prime-R_1)/(2\sigma)$, we have the following
theorem which will be proved in Section~\ref{sec:proof}.
\begin{theorem}[Mutual Information Spectrum] \label{thm:main1}
For ReRAM channel
$\left\{W^\Phi_N(\cdot|\cdot)\right\}_{N=1}^\infty$ with input
distribution i.i.d. Bernoulli $(q)$,
\begin{eqnarray}
\lim_{N\rightarrow\infty}\mathcal{I}_q^N\overset{p}{=}&&\!\!\!\!\!\!\!\!\!\!
\mathcal{I}_q^\infty \label{eq:Iqinf} \\
\textrm{with} \ \  P_{\mathcal{I}_q^\infty}(x)
=&&\!\!\!\!\!\!\!\!\!\!\sum_{k=0}^Kp_k\sum_{k^\prime=0}^k
\binom{k}{k^\prime}q^{k^\prime}(1-q)^{k-k^\prime}
\delta\left[x-C_q(\gamma^\prime)\right.\nonumber\\
&& \ \ \ \  \left.-(1-q^2)^{k^\prime}\left(C_q(\gamma)-
C_q(\gamma^\prime)\right)\right]\label{eq:PI}.
\end{eqnarray}
\end{theorem}

Theorem~\ref{thm:main1} indicates that the mutual information
spectrum of ReRAM channel has nonzero probability only at $K+1$
discrete points
$C_q(\gamma^\prime)+(1-q^2)^{k^\prime}(C_q(\gamma)-C_q(\gamma^\prime)),
k^\prime=0, 1, ..., K$, which means that the channel has $K+1$
statuses corresponding to the $K+1$ types of SF pattern. The proof
of Theorem~\ref{thm:main1} involves an analysis of the
\textit{asymptotic SP rate}  (Lemma~\ref{lem:SPrate},
Sec.~\ref{sec:proof}) showing that fixing $k^\prime$ active SFs in
the array, as $N\rightarrow\infty$, the fraction of cells affected
by sneak paths is about $1-(1-q^2)^{k^\prime}$. Intuitively, a
cell without sneak-path interference can store information with
maximum rate of $C_q(\gamma)$ and that with sneak-path
interference can only store information with maximum rate of
$C_q(\gamma^\prime)$. Thus, the overall maximum achievable
information rate is
$C_q(\gamma^\prime)+(1-q^2)^{k^\prime}(C_q(\gamma)-C_q(\gamma^\prime))$.
The result of (\ref{eq:PI}) is formed by the probability-weighted
sum of the maximum achievable information rate with respect to the
probability distribution of the active SF number $k^\prime$.

Theorem~\ref{thm:main1} immediately leads to the following
theorem.
\begin{theorem}[Single-Array Coding Rate]\label{thm:main2}
The maximum achievable rate of ReRAM channel
$\left\{W^\Phi_N(\cdot|\cdot)\right\}_{N=1}^\infty$  with
arbitrary small error probability is
\begin{eqnarray}
\mathcal{R}=\sup_q\left[C_q(\gamma^\prime)+(1-q^2)^K
\left(C_q(\gamma)-C_q(\gamma^\prime)\right)\right].\label{eq:R}
\end{eqnarray}
\end{theorem}

\emph{Proof:} Since $C_q(\gamma)>C_q(\gamma^\prime)$,
$C_q(\gamma^\prime)+(1-q^2)^{k^\prime}(C_q(\gamma)-C_q(\gamma^\prime))$
decreases with $k^\prime$. Therefore, the spectral inf-mutual
information rate $\underline{\mathcal{I}}_q=C_q(\gamma^\prime)
+(1-q^2)^K(C_q(\gamma)-C_q(\gamma^\prime))$. Using
Definition~\ref{def:R}, we have (\ref{eq:R}). \myQED

Theorem~\ref{thm:main2} reveals the following facts of
single-array coding:
\begin{itemize}

\item The maximum achievable rate is subject to the worst case of
the channel status, which is determined by the maximum possible
number of SFs in an array.

\item For $K=0$, $\mathcal{R}=C_{0.5}(\gamma)$ is the capacity of
a binary-input additive Gaussian channel without sneak path
interference.

\item As $K\rightarrow\infty$, $\mathcal{R}\rightarrow
C_{0.5}(\gamma^\prime)$ is the capacity of a binary-input additive
Gaussian channel when all the memory cells (with HRS) are affected
by sneak paths.

\item For finite $K>0$,
$C_{0.5}(\gamma^\prime)<\mathcal{R}<C_{0.5}(\gamma)$ and
$\mathcal{R}$ decreases as the maximum number of SFs $K$
increases.
\end{itemize}

The cask effect of single-array coding is due to the randomness of
the SFs in the array. To improve the achievable rate, we consider
an across-array coding strategy, where data of $T$ arrays is
jointly encoded as one codeword. The fundamental difference
between the single- and across-array coding is that, in the
single-array coding, the codeword experiences only one ReRAM
channel, while in the across $T$-array coding, a codeword
experiences $T$ independent ReRAM channels. In the following, we
show that when $T$ is large, across-array coding achieves an
ergodic rate, which is the expectation of (\ref{eq:Iqinf}).
\begin{theorem}[Across-Array Coding Rate]\label{thm:main3}
For ReRAM channel
$\left\{W^\Phi_N(\cdot|\cdot)\right\}_{N=1}^\infty$ with across
$T$-array coding, the maximum achievable rate as
$T\rightarrow\infty$ is
\begin{eqnarray}
\mathcal{R}=\sup_q\left[C_q(\gamma^\prime)+\left(C_q(\gamma)
-C_q(\gamma^\prime)\right)\sum_{k=0}^Kp_k(1-q^3)^k\right].\label{eq:Rc}
\end{eqnarray}
\end{theorem}
\emph{Proof:} Since the SF patterns of the $T$ arrays  are
independent, the $T$-array channel consists of $T$ independent
channels $\left\{W^{\Phi^t}_N(\cdot|\cdot)\right\}_{N=1}^\infty$
with inputs $X^{N\times N}_t$ and outputs $Y^{N\times N}_t, t=1,
2,...,T$. The corresponding mutual information density rate of
this $T$-array channel is
\begin{eqnarray}
\lim_{N\rightarrow\infty}\mathcal{I}_q^{N,T}&&\!\!\!\!\!\!\!\!\!\!
=\lim_{N\rightarrow\infty}\frac{1}{TN^2}\log\frac{\prod_{t=1}^TW^{\Phi^t}_N
\left(Y^{N\times N}_t|X^{N\times
N}_t\right)}{\prod_{t=1}^TP_{Y^{N\times N}}
\left(Y^{N\times N}_t\right)}\nonumber\\
&&\!\!\!\!\!\!\!\!\!\!\overset{p}{=}\frac{1}{T}\sum_{t=1}^T
\mathcal{I}_q^{t,\infty},
\end{eqnarray}
where $\mathcal{I}_q^{t,\infty}, t=1, 2,...,T$, are i.i.d. with
probability density function $P_{\mathcal{I}_q^\infty}(x)$ as
shown in (\ref{eq:PI}). The law of large numbers leads to
\begin{eqnarray}
\!\!\!\!\!\!\!\!\!\!\!\lim_{T\rightarrow\infty}\frac{1}{T}
\sum_{t=1}^T\mathcal{I}_q^{t,\infty}&&\!\!\!\!\!\!\!\!\!
\overset{p}{=}\int_{-\infty}^{\infty}
xP_{\mathcal{I}_q^\infty}(x)dx \nonumber\\
&&\!\!\!\!\!\!\!\!\!\!\!=C_q(\gamma^\prime)+\left(C_q(\gamma)
-C_q(\gamma^\prime)\right)\sum_{k=0}^Kp_k(1-q^3)^k\label{eq:Rcq}
\end{eqnarray}
where we used $\int_{-\infty}^{\infty}
x\delta\left[x-a\right]dx=a$ and
$\sum_{i=0}^k\binom{k}{i}a^ib^{k-i}=(a+b)^k$. Since the right hand
side of  (\ref{eq:Rcq}) is a constant, it is exactly the spectral
inf-mutual information rate $\underline{\mathcal{I}}_q$.
Optimizing $\underline{\mathcal{I}}_q$ with respect to $q$, we
obtain (\ref{eq:Rc}). \myQED

Since for $K>0$, $\sum_{k=0}^Kp_k(1-q^3)^k>(1-q^2)^K$ always
holds, across-array coding  achieves a higher rate than
single-array coding, which coincides with the simulation results
observed in coded ReRAM system \cite{SongTcom}. This gain is due
to the channel ``diversity" obtained by assigning coded bits to
multiple arrays, and the channel status is averaged. In summary,
Theorem~\ref{thm:main3} reveals:

\begin{itemize}
\item Across-array coding  achieves a higher rate than
single-array coding.

\item Across-array coding achieves the expected value of the
mutual information rate, which is not subject to the worse channel
status.
\end{itemize}

\begin{figure}[t]
\includegraphics[width=
3.5 in]{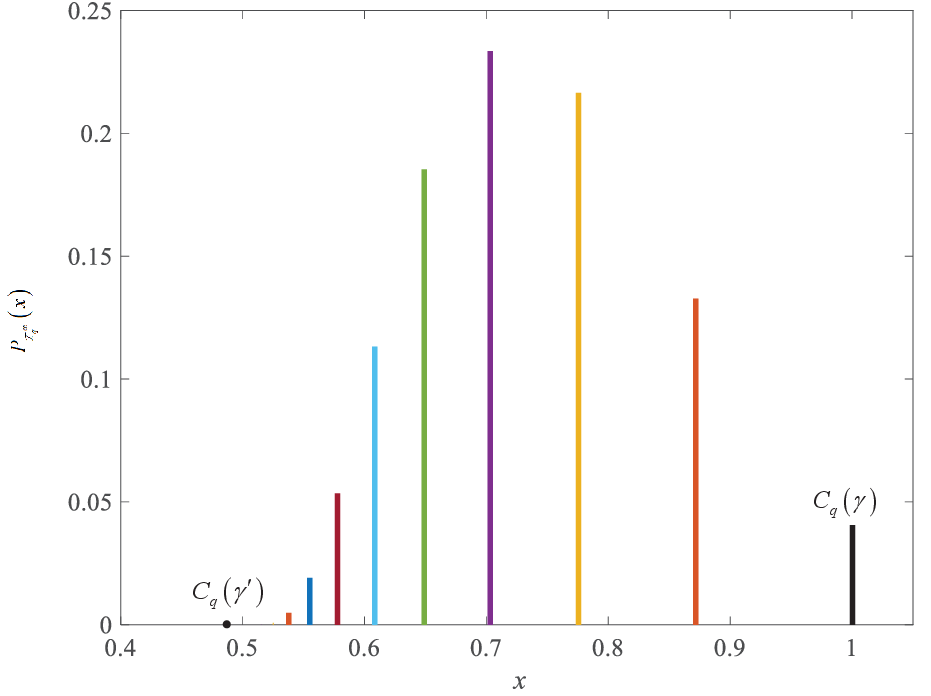} \centering \caption{Mutual information spectrum
of ReRAM channel with $q=0.5$, $R_1=100\ \Omega, R_0=1000\ \Omega,
R_s=250\ \Omega, \sigma=50$. SF number is with
$\mathcal{B}_K(n,\mu)$ with $n=256\times 256, \mu=10^{-4}, K=8$.}
\label{fig:Spectrum}
\end{figure}
\begin{figure}[t]
\includegraphics[width=
3.5 in]{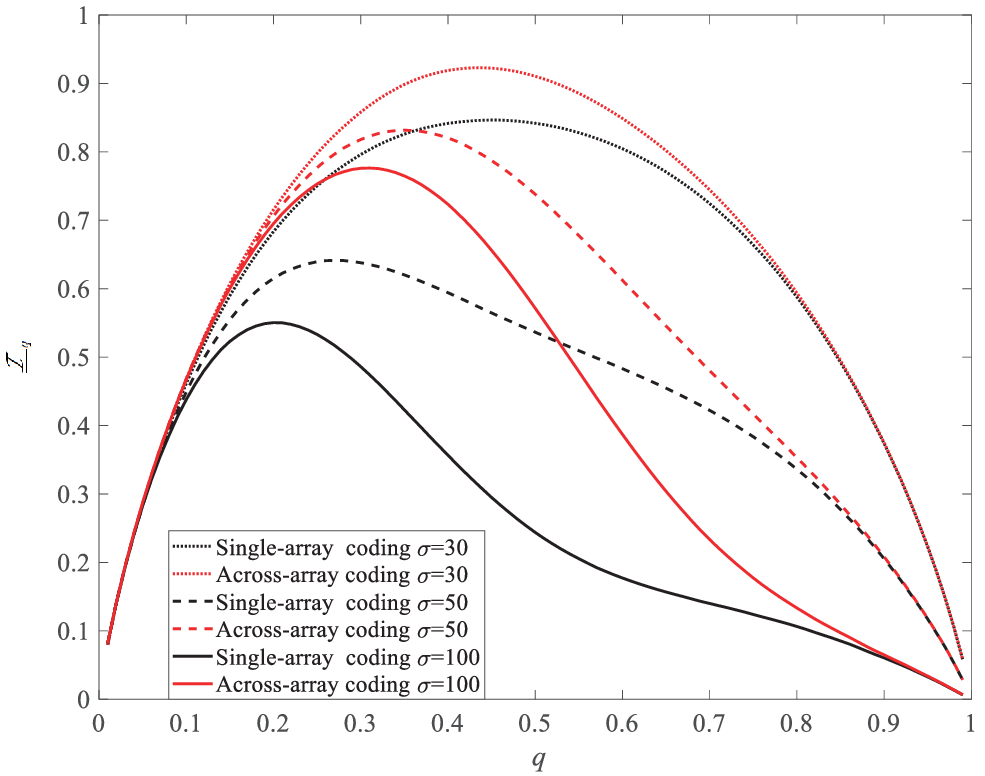} \centering \caption{Spectral inf-mutual
information rate of ReRAM channel for both single- and
across-array codings with varying $q$. $R_1=100\ \Omega, R_0=1000\
\Omega, R_s=250\ \Omega, K=8$.} \label{fig:Capacity_q}
\end{figure}

\begin{figure}[t]
\includegraphics[width=
3.5 in]{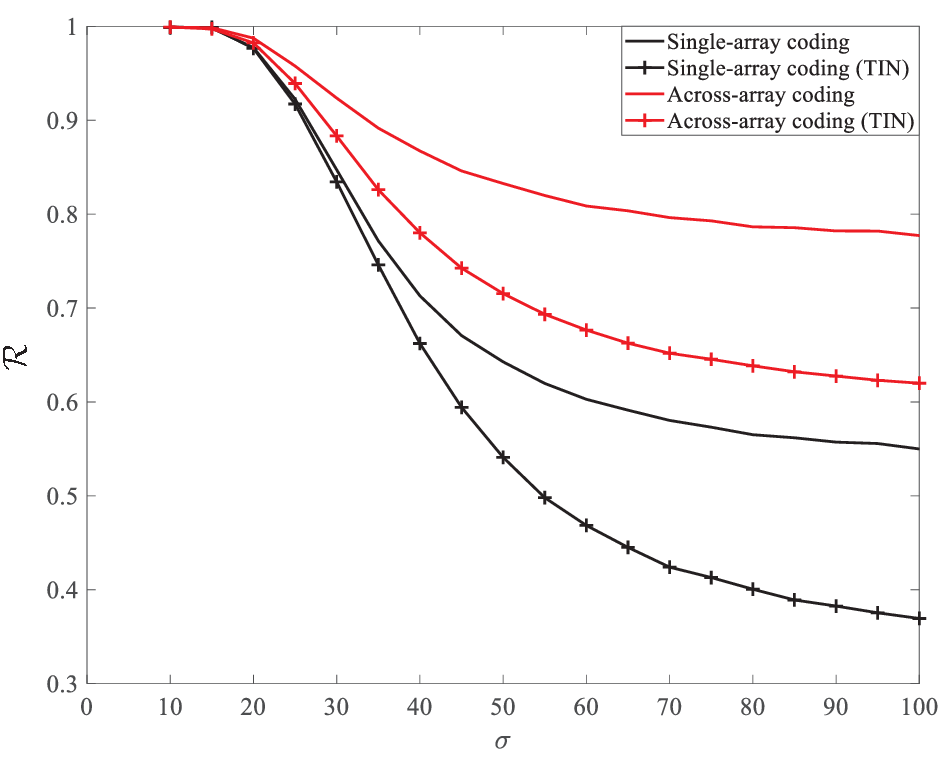} \centering \caption{Maximum achievable
rate $\mathcal{R}$ of ReRAM channel under both of the optimal and
TIN decodings. $R_1=100\ \Omega, R_0=1000\  \Omega, R_s=250\
\Omega, K=8$. } \label{fig:Capacity_sigma}
\end{figure}

We provide some numerical results for Theorems~\ref{thm:main1}
-\ref{thm:main3} in
Figs.~\ref{fig:Spectrum}-\ref{fig:Capacity_sigma}. We assume that
the SF number in each array is with a truncated binomial
distribution $\mathcal{B}_K(n,\mu)$ with $n=256\times 256,
\mu=10^{-4}, K=8$, which is obtained by bounding a binomial
distributed variable by a maximum value of $K$. Hence
$\mathcal{B}_K(n,\mu)$ is an approximation of binomial
distribution $\mathcal{B}(n,\mu)$  used in
\cite{Ben,CZH,SongTcom}. According to the definition of binomial
distribution $\mathcal{B}(n,\mu)$, $n$ and $\mu$ denote the total
number of Bernoulli trials and the successful probability of each
trial, respectively. Here, $n$ plays the role of the total number
of memory cells of an array, and $\mu$ plays the role of the SF
probability of each cell. The values of $n$ and $\mu$ used here
capture the features of large memory array with low SF probability
in ReRAM. We follow the parameter settings of
\cite{Ben,CZH,SongTcom} with $R_1=100\ \Omega, R_0=1000\  \Omega,
R_s=250\ \Omega$, and hence
$\gamma=(R_0-R_1)/(2\sigma)=450/\sigma$ and
$\gamma^\prime=(R_0^\prime-R_1)/(2\sigma)=50/\sigma$.
Fig.~\ref{fig:Spectrum} illustrates the mutual information
spectrum $P_{\mathcal{I}_q^\infty}$ derived in
Theorem~\ref{thm:main1} for $q=0.5, \sigma=50$. In this case,
$C_q(\gamma)\approx1$ and $C_q(\gamma^\prime)\approx0.4861$. The
spectrum has non-zero values at
$\mathcal{I}_q^\infty=C_q(\gamma^\prime)+(1-q^2)^k(C_q(\gamma)
-C_q(\gamma^\prime)), k=0, 1, ..., 8$, corresponding to the $9$
statuses of the ReRAM channel.

Fig.~\ref{fig:Capacity_q} shows the spectral inf-mutual
information rate $\underline{\mathcal{I}}_q$ of ReRAM channel for
both single- and across-array codings with varying $q$ obtained in
Theorem~\ref{thm:main2} and Theorem~\ref{thm:main3}. Across-array
coding achieves much higher rate than single-array coding due to
the diversity gain.  Moreover, the maximum rates are achieved when
$q<0.5$ for both single- and across-array codings, i.e., uniform
input distribution is not optimal. It is surprising that for ReRAM
channel the capacity loss by using uniform distribution is
significant. For example, for $\sigma=100$, the maximum achievable
rates of single- and across-array codings are $0.55$ bits/cell and
$0.7778$ bits/cell at $q=0.2$ and $q=0.31$, respectively. If we
use the uniform input distribution with $q=0.5$ instead, the
maximum achievable rates are only $0.2448$ bits/cell and $0.5723$
bits/cell, leading to a capacity loss of about $55\%$ and $26\%$,
respectively.  This is quite different from the results  observed
from conventional i.i.d. asymmetric channels without memory, such
as the binary asymmetric channel (BAC) and $Z$-channel, for which,
the capacity loss by using uniform input distribution is less than
$6\%$ \cite{Shulman,Bennatan}. The main reason that causes ReRAM
channel different from conventional asymmetric channels is that
the sneak path occurrence is greatly affected by the channel input
distribution, which can be seen from the analysis in
Section~\ref{sec:proof}.    For the same reason, it is desirable
to shape input data into the optimal distribution before writing
it into the memory array. A practical data shaping with the
corresponding de-shaping scheme was proposed in \cite{SongTcom}
and its performance gain over ReRAM channel was confirmed in
\cite{SongTcom}.

Fig.~\ref{fig:Capacity_sigma} shows the maximum achievable rates
$\mathcal{R}$ of ReRAM channel with the optimal $q$ for different
$\sigma$. For comparison, we also illustrate the maximum
achievable rate of ReRAM channel under the TIN detection/decoding
employed in \cite{Ben,CZH,SongTcom}. The TIN detection ignores the
channel correlation by treating the sneaking path interference as
noise and its capacity is derived as an memoryless channel
\cite{SongTcom}. Since the TIN decoding is suboptimal, its maximum
achievable rate is much lower than the achievable rate derived in
this paper. This capacity loss is insignificant when the noise
level is low, but increases as the noise level increases. For
$\sigma=100$, TIN decoding achieves a maximum rate of $0.37$ and
$0.6189$  bits/cell with single- and across-array codings, which
decrease the channel capacity by $33\%$ and $20\%$, respectively.
Therefore, a more sophisticated data detection/decoding scheme
with a joint data and sneak path processing is desirable if the
corresponding complexity is affordable to the system. One of such
detection schemes is proposed in our recent work \cite{SongArxiv}.

\section{Mutual Information Spectrum Analysis\\ (Proof of
Theorem~\ref{thm:main1})} \label{sec:proof} For a general ReRAM
channel defined by Definition~\ref{def:ReRAM}, due to the channel
memory, the mutual information spectrum cannot be derived directly
using the law of large numbers as in Example~\ref{eg:iid}. In this
section, we present a mutual information spectrum analysis for the
general ReRAM channel, which serves as the proof of
Theorem~\ref{thm:main1}.

In the following, we always assume that  $X^{N\times N}$ and
$Y^{N\times N}$ are the input and output of  ReRAM channel
$W^\Phi_N(\cdot|\cdot)$ and entries of $X^{N\times N}$ are with
i.i.d. Bernoulli $(q)$.  We split $X^{N\times N}$ into two parts
$X_{sf}$ and $X_{sf}^c$, where $X_{sf}$ contains the entries of SF
rows and columns and $X_{sf}^c$ contains all the remaining
entries. We first need the following lemma.
\begin{lemma}\label{lem:replace}
 For any given $\epsilon>0$, we have
\begin{eqnarray}
\lim_{N\rightarrow\infty}\textrm{Pr}\left(\frac{1}{N^2}
\left|\log\frac{P_{Y^{N\times N}} \left(Y^{N\times
N}\right)}{P_{Y^{N\times N}|X_{sf},\Phi} \left(Y^{N\times
N}|X_{sf},\Phi\right)}\right|<\frac{\epsilon}{2}\right)
&&\!\!\!\!\!\!\!\!\!\!=1,\nonumber\\
\lim_{N\rightarrow\infty}\textrm{Pr}\left(\frac{1}{N^2}
\left|\log\frac{P_{Y^{N\times N}|X^{N\times N}} \left(Y^{N\times
N}|X^{N\times N}\right)} {P_{Y^{N\times N}|X^{N\times
N},\Phi}\left(Y^{N\times N}|X^{N\times N},
\Phi\right)}\right|<\frac{\epsilon}{2}\right)&&\!\!\!\!\!\!\!\!\!\!
=1\nonumber.
\end{eqnarray}
\end{lemma}

\emph{Proof:} We prove the first equation and the second one is
proved in Appendix~\ref{lem1proof2}. Let $\Psi$ be the set that
includes all the possible realizations of $(X_{sf},\Phi)$. Define
the following sets for $(x_{sf},\varphi)\in \Psi, y^{N\times N}\in
\mathbb{R}^{N\times N}$,
\begin{eqnarray}
A_N&&\!\!\!\!\!\!\!\!\!\!=\left\{\left(x_{sf},\varphi, y^{N\times
N}\right)\left|\log\frac{P_{Y^{N\times N}} \left(y^{N\times
N}\right)}{P_{Y^{N\times N}|X_{sf},\Phi} \left(y^{N\times
N}|x_{sf},\varphi\right)}\leq\frac{-\epsilon N^2}{2}
\right.\right\},\nonumber\\
B_N&&\!\!\!\!\!\!\!\!\!\!=\left\{\left(x_{sf},\varphi, y^{N\times
N}\right)\left|\log\frac{P_{Y^{N\times N}} \left(y^{N\times
N}\right)}{P_{Y^{N\times N}|X_{sf},\Phi} \left(y^{N\times
N}|x_{sf},\varphi\right)}\geq\frac{\epsilon N^2}{2}
\right.\right\}\nonumber.
\end{eqnarray}
Since
\begin{eqnarray}
&&\!\!\!\!\!\!\!\!\!\!\textrm{Pr}\left(\frac{1}{N^2}
\left|\log\frac{P_{Y^{N\times N}} \left(Y^{N\times
N}\right)}{P_{Y^{N\times N}|X_{sf},\Phi} \left(Y^{N\times
N}|X_{sf},\Phi\right)}\right|<\frac{\epsilon}{2}\right)
\nonumber\\ =&&\!\!\!\!\!\!\!\!\!\!1-\sum_{(x_{sf},\varphi,
y^{N\times N}) \in A_N\cup B_N}P_{X_{sf},\Phi,Y^{N\times
N}}\left(x_{sf},\varphi, y^{N\times N}\right),\nonumber
\end{eqnarray}
to prove the lemma, we only need to show that
\begin{equation}
\sum_{(x_{sf},\varphi, y^{N\times N})\in A_N\cup
B_N}P_{X_{sf},\Phi, Y^{N\times N}}\left(x_{sf},\varphi,y^{N\times
N}\right)\rightarrow0. \nonumber
\end{equation}

Since a $k$-SF pattern $\varphi$ leads to maximum of $2kN-k^2$
entries in $x_{sf}$ (with $k$ SF rows and $k$ SF columns), which
happens when $\varphi$ is a scattered $k$-SF pattern
(Definition~\ref{def:scattered}), we have
$|\Psi|<\sum_{k=0}^K2^{2kN-k^2}\binom{N^2}{k}<(K+1)4^{KN}N^{2K}$.
Using the fact that the joint probability is smaller than or equal
to the marginal probability, we have
\begin{eqnarray}
&&\!\!\!\!\!\!\!\!\!\!\sum_{(x_{sf},\varphi, y^{N\times N})\in
A_N} P_{X_{sf},\Phi,Y^{N\times N}}\left(x_{sf},\varphi,y^{N\times
N}\right)
\nonumber\\
\leq&&\!\!\!\!\!\!\!\!\!\! \sum_{(x_{sf},\varphi, y^{N\times
N})\in A_N}
P_{Y^{N\times N}}\left(y^{N\times N}\right)\nonumber\\
\leq&&\!\!\!\!\!\!\!\!\!\! e^{-\epsilon
N^2/2}\sum_{(x_{sf},\varphi, y^{N\times N})\in A_N}P_{Y^{N\times
N}|X_{sf},\Phi}\left(y^{N\times N}
|x_{sf},\varphi\right)\nonumber\\
\leq&&\!\!\!\!\!\!\!\!\!\! e^{-\epsilon
N^2/2}\sum_{(x_{sf},\varphi)\in \Psi}\sum_{y^{N\times
N}\in\mathbb{R}^{N\times N}}P_{Y^{N\times N}|X_{sf},
\Phi}\left(y^{N\times N}|x_{sf},\varphi\right)\nonumber\\
<&&\!\!\!\!\!\!\!\!\!\!e^{-\epsilon N^2/2}
(K+1)4^{KN}N^{2K}\rightarrow 0, \nonumber
\end{eqnarray}
where the second inequality is obtained directly from the
definition of $A_N$, and in the last inequality, we used the fact
that for any pair of $(x_{sf},\varphi)$, $\sum_{y^{N\times N}\in
\in\mathbb{R}^{N\times N}}P_{Y^{N\times
N}|X_{sf},\Phi}\left(y^{N\times N}|x_{sf},\varphi\right)\!=\!1$.

Moreover,
\begin{eqnarray}
&&\!\!\!\!\!\!\!\!\!\!\sum_{(x_{sf},\varphi, y^{N\times N}) \in
B_N}P_{X_{sf},\Phi,Y^{N\times N}}\left(x_{sf},\varphi,
y^{N\times N}\right)\nonumber\\
=&&\!\!\!\!\!\!\!\!\!\! \sum_{(x_{sf},\varphi, y^{N\times N}) \in
B_N}P_{X_{sf},\Phi}\left(x_{sf},\varphi\right)P_{Y^{N\times N}
|X_{sf},\Phi}\left(y^{N\times N}|x_{sf},\varphi\right)\nonumber\\
\leq&&\!\!\!\!\!\!\!\!\!\! e^{-\epsilon
N^2/2}\sum_{(x_{sf},\varphi, y^{N\times N})\in
B_N}P_{X_{sf},\Phi}\left(x_{sf},\varphi\right)
P_{Y^{N\times N}}\left(y^{N\times N}\right)\nonumber\\
\leq&&\!\!\!\!\!\!\!\!\!\! e^{-\epsilon
N^2/2}\sum_{(x_{sf},\varphi)
\in\Psi}P_{X_{sf},\Phi}\left(x_{sf},\varphi\right)\sum_{y^{N\times
N} \in\mathbb{R}^{N\times N}}P_{Y^{N\times N}}\left(y^{N\times
N}\right)
\nonumber\\
=&&\!\!\!\!\!\!\!\!\!\! e^{-\epsilon N^2/2}\rightarrow 0,\nonumber
\end{eqnarray}
where we used equations
$\sum_{(x_{sf},\varphi)\in\Psi}P_{X_{sf},\Phi}
\left(x_{sf},\varphi\right)=1$ and $\sum_{y^{N\times N}
\in\mathbb{R}^{N\times N}}P_{Y^{N\times N}}\left(y^{N\times
N}\right)=1$. The lemma is proved. \myQED

An explanation of Lemma~\ref{lem:replace}  is that since the
maximum SF number $K$ is small compared to the array size $N$,
with or without the knowledge of SF pattern $\Phi$ and SF rows and
columns $X_{sf}$ does not affect greatly the joint probability of
the readback signal array $Y^{N\times N}$.

Let
\begin{equation}\label{eq:Iq1}
\widetilde{\mathcal{I}}_q^N=\frac{1}{N^2}\log\frac{P_{Y^{N\times
N} |X^{N\times N},\Phi}\left(Y^{N\times N}|X^{N\times
N},\Phi\right)}{P_{Y^{N\times N}|X_{sf},\Phi}\left(Y^{N\times
N}|X_{sf},\Phi\right)}.
\end{equation}
Definition~\ref{def:I} and Lemma~\ref{lem:replace} lead to
\begin{corollary}\label{cor:Iq} For any $\epsilon>0$,
\begin{equation}
\lim_{N\rightarrow\infty}\textrm{Pr}\left(\left|\mathcal{I}_q^N
-\widetilde{\mathcal{I}}_q^N\right|<\epsilon\right)=1.
\end{equation}
\end{corollary}
Therefore, we can derive the mutual information spectrum based on
$\widetilde{\mathcal{I}}_q^N$ instead.

To proceed with the analysis, we further need the following two
lemmas, which show the asymptotic behaviours of the SF pattern and
sneak path occurrence rate. First, we show that as
$N\rightarrow\infty$, a randomly chosen SF pattern is a scattered
SF pattern with probability 1. Let $\overline{\Omega}_N^k$ denote
the set that includes all the scattered $k$-SF patterns and
$\overline{\Omega}_N$ be the universal set that includes all the
scattered SF patterns. We have the following lemma.
\begin{lemma} [Converge to Scattered SF Pattern]\label{lem:scattered}
\begin{eqnarray}
\lim_{N\rightarrow\infty}\textrm{Pr}\left(\Phi\in\overline{\Omega}_N\right)
&&\!\!\!\!\!\!\!\!\!\!=1, \label{eq:lemeq1}\\
\lim_{N\rightarrow\infty}\textrm{Pr}\left(\Phi\in\overline{\Omega}_N^k\right)
&&\!\!\!\!\!\!\!\!\!\!=p_k,\ k=0, 1,..., K. \label{eq:pk}
\end{eqnarray}
\end{lemma}
\emph{Proof:} Since $|\overline{\Omega}_N^k|=\binom{N}{k}^2k!$,
according to the probability distribution of SF patterns defined
in Section~\ref{sec:main results}, we have
\begin{eqnarray}
1&&\!\!\!\!\!\!\!\!\!\!\geq\lim_{N\rightarrow\infty}\textrm{Pr}
\left(\Phi\in\overline{\Omega}_N\right)\\
&&\!\!\!\!\!\!\!\!\!\!=\lim_{N\rightarrow\infty}\sum_{k=1}^K\textrm{Pr}
\left(\Phi\in\overline{\Omega}_N^k\right)\\
&&\!\!\!\!\!\!\!\!\!\!=\lim_{N\rightarrow\infty}\sum_{k=1}^K
p_kk!{\binom{N}{k}^2}\left/{\binom{N^2}{k}}\right.\\
&&\!\!\!\!\!\!\!\!\!\!=\lim_{N\rightarrow\infty}\sum_{k=1}^Kp_k
\frac{N^2(N-1)^2\cdots(N-k+1)^2}{N^2(N^2-1)\cdots(N^2-k+1)}\\
&&\!\!\!\!\!\!\!\!\!\!\geq\lim_{N\rightarrow\infty}\sum_{k=1}^Kp_k
\left(\frac{N-k+1}{N}\right)^{2k}\\
&&\!\!\!\!\!\!\!\!\!\!\geq \lim_{N\rightarrow\infty}
\left(\frac{N-K+1}{N}\right)^{2K}=1. \label{eq:final}
\end{eqnarray}

As (\ref{eq:lemeq1}) leads to
$\lim_{N\rightarrow\infty}\textrm{Pr}
\left(\Phi\notin\overline{\Omega}_N\right)=0$, we have
\begin{eqnarray}
p_k\geq&&\!\!\!\!\!\!\!\!\!\!\lim_{N\rightarrow\infty}
\textrm{Pr}\left(\Phi\in\overline{\Omega}_N^k\right)\nonumber\\
\geq&&\!\!\!\!\!\!\!\!\!\! \lim_{N\rightarrow\infty}
\left(\textrm{Pr}\left(\Phi\in\Omega_N^k\right)-\textrm{Pr}
\left(\Phi\notin\overline{\Omega}_N\right)\right)=p_k\nonumber.
\end{eqnarray}
The lemma is proved. \myQED

\begin{definition}
SP rate $\alpha_{sp}^N(\varphi)$ of memory array $X^{N\times N}$
with given SF pattern $\varphi$ is
\begin{equation}
\alpha_{sp}^N(\varphi)=\frac{1}{N^2}\sum_{m=1}^N\sum_{n=1}^N
\bigcup_{(i,j)\in\varphi}X_{m,j}X_{i,j}X_{i,n}.\label{eq:SPrate}
\end{equation}
\end{definition}
Since if $\bigcup_{(i,j)\in\varphi}X_{m,j}X_{i,j}X_{i,n}=1$, cell
$(m,n)$ will potentially be an SP cell and if
$\bigcup_{(i,j)\in\varphi}X_{m,j}X_{i,j}X_{i,n}=0$, cell $(m,n)$
will never be an SP cell. Therefore, SP rate is the fraction of
cells that are potential to be affected by sneak paths in an
array.

The following lemma shows that if $\varphi$ is a scattered SF
pattern, $\alpha_{sp}^N(\varphi)$  converges in probability to
some determined values which are  related to the number of SFs and
$q$. For SF pattern $\varphi$, denote $\varphi_1$ and $\varphi_2$
as the SF row and column index sets, respectively. For example,
for $\varphi=\{(1,2), (3,4)\}$,  $\varphi_1=\{1, 3\}$ and
$\varphi_2=\{2, 4\}$.
\begin{lemma} [Asymptotic SP Rate] \label{lem:SPrate}
For any $\varphi\in \overline{\Omega}_N^k$ and $\epsilon>0$, it
holds that
\begin{eqnarray}
&&\!\!\!\!\!\! \underset{N\rightarrow\infty}\lim
\alpha_{sp}^N(\varphi)
\overset{p}{=}\alpha_k  \\
\textrm{with} &&\!\!\!\!\!\!
P_{\alpha_k}(x)=\sum_{k^\prime=0}^k\binom{k}{k^\prime}
q^{k^\prime}(1-q)^{k-k^\prime}\delta\left[x-1+(1-q^2)^{k^\prime}
\right]\nonumber.
\end{eqnarray}
\end{lemma}

\emph{Proof:} Let $\varphi^*=\{(i, j)| (i, j)\in\varphi\
\textrm{and} \ X_{i,j}=1\}$ be the set that includes all the
locations of active SFs in $\varphi$. Since $|\varphi|=k$, we have
$\textrm{Pr}\left(|\varphi^*|=k^\prime\right)=\binom{k}{k^\prime}
q^{k^\prime}(1-q)^{k-k^\prime}$ for $k^\prime=0, 1,...,k$. To
prove the lemma, we only need to show that for any given
$\epsilon>0$,
$\lim_{N\rightarrow\infty}\textrm{Pr}\left(\left.|\alpha_{sp}^N(\varphi)
-1+(1-q^2)^{k^\prime}|<\epsilon\right|
|\varphi^*|=k^\prime\right)=1$.

Since $\bigcup_{(i,j)\in\varphi}X_{m,j}X_{i,j}X_{i,n}
=\bigcup_{(i,j)\in\varphi^*}X_{m,j}X_{i,n}\in\{0, 1\}$,
$|\varphi_1|=|\varphi_2|=k$, and deleting $2Nk-k^2 = o(N^2)$ terms
in (\ref{eq:SPrate}) does not affect its limitation, we have
\begin{eqnarray}
\lim_{N\rightarrow\infty}\alpha_{sp}^N(\varphi)&&\!\!\!\!\!\!\!\!\!\!
=\lim_{N\rightarrow\infty}\frac{1}{(N-k)^2}\sum_{m\notin\varphi_1}
\sum_{n\notin\varphi_2}\bigcup_{(i,j)\in\varphi^*}X_{m,j}X_{i,n}\nonumber\\
&&\!\!\!\!\!\!\!\!\!\!=\lim_{N\rightarrow\infty}\frac{1}{(N-k)^2}
\sum_{m\notin\varphi_1}\sum_{n\notin\varphi_2}\left(1-\prod_{(i,j)
\in\varphi^*}(1-X_{m,j}X_{i,n})\right)\nonumber\\
&&\!\!\!\!\!\!\!\!\!\!=1-\lim_{N\rightarrow\infty}\frac{1}{N-k}
\sum_{m\notin\varphi_1}\frac{1}{N-k}\sum_{
n\notin\varphi_2}A_{m,n}, \nonumber
\end{eqnarray}
where $A_{m,n}=\prod_{(i,j)\in\varphi^*}(1-X_{m,j}X_{i,n})$. Let
\begin{equation}
\tilde{\alpha}_{sp}^N(\varphi)\triangleq1-\frac{1}{N-k}
\sum_{m\notin\varphi_1}\frac{1}{N-k}\sum_{ n\notin\varphi_2}
A_{m,n}.\label{eq:alpha_tilde}
\end{equation}
We have $\lim_{N\rightarrow\infty}\alpha_{sp}^N(\varphi)
=\lim_{N\rightarrow\infty}\tilde{\alpha}_{sp}^N(\varphi)$.

Since $A_{m,n}, m\notin\varphi_1, n\notin\varphi_2$, are not
independent, we cannot apply the law of large numbers directly to
(\ref{eq:alpha_tilde}). In the following, we prove the convergence
of (\ref{eq:alpha_tilde}) in two steps. We first bound the inner
summation $\frac{1}{N-k}\sum_{n\notin\varphi_2}A_{m,n}$ using
large deviation theory based on types. The outer summation
$\frac{1}{N-k}\sum_{m\notin\varphi_1}$ is then bounded using the
weak law of large numbers and union bound.

Given $X_{m,j}, m\notin\varphi_1,  j\in\varphi^*_2$, $A_{m,n},
n\notin\varphi_2$, are i.i.d. Bernoulli random variables with mean
\begin{eqnarray}
\mu_m&&\!\!\!\!\!\!\!\!\!\!=\textrm{Pr}(A_{m,n}=1)\nonumber\\
&&\!\!\!\!\!\!\!\!\!\!=\textrm{E}\left[\prod_{(i,j)\in\varphi^*}
(1-X_{m,j}X_{i,n})\right]\nonumber\\
&&\!\!\!\!\!\!\!\!\!\!
=\prod_{(i,j)\in\varphi^*}(1-X_{m,j}\textrm{E}
\left[X_{i,n}\right])\nonumber\\
&&\!\!\!\!\!\!\!\!\!\! =\prod_{j\in\varphi^*_2}(1-X_{m,j}q)
\end{eqnarray}
and $\textrm{Pr}(A_{m,n}=0)=1-\mu_m$.

Let $\mathbb{E}_m$ be the event of $\sum_{n\notin\varphi_2}A_{m,n}
\in
\mathcal{A}_{m}=\left\{d\left||\frac{d}{N-k}-\mu_m|\right.<\epsilon_1,
d=0, 1, ..., N-k\right\}$ and $\mathbb{E}_m^c$ be the event of
$\sum_{n\notin\varphi_2}A_{m,n} \in
\mathcal{A}_{m}^c=\left\{d\left|
|\frac{d}{N-k}-\mu_m|\right.\geq\epsilon_1, d=0, 1, ...,
N-k\right\}$. Let $\mathcal{P}_p=(p, 1-p)$ denote the Bernoulli
$(p)$ distribution. Since $\sum_{n\notin\varphi_2}A_{m,n}$ is a
binomial distributed variable with
$\sum_{n\notin\varphi_2}A_{m,n}\sim \mathcal{B}(N-k, \mu_m)$, we
have
\begin{eqnarray}
\textrm{Pr}\left(\mathbb{E}_m^c \right)&&\!\!\!\!\!\!\!\!\!\!
=\sum_{d\in\mathcal{A}_m^c}\binom{N-k}{d}\mu_m^d(1-\mu_m)^{N-k-d}
\label{eq:PEm1}\\
&&\!\!\!\!\!\!\!\!\!\!=\sum_{d\in\mathcal{A}_m^c}
\binom{N-k}{d}2^{-(N-k)\left(H\left(\frac{d}{N-k}\right)+D_{KL}
\left(\mathcal{P}_{\frac{d}{N-k}}\Vert\mathcal{P}_{\mu_m}\right)\right)},
\label{eq:PEm2}
\end{eqnarray}
where $H(x)=-x\log_2x-(1-x)\log_2(1-x)$ is the binary entropy
function and $D_{KL}(\cdot\Vert\cdot)$ is the Kullback-Leibler
divergence of two probability distributions.

Since $|\mathcal{A}_m^c|\leq N-k+1, \binom{N-k}{d} \leq
2^{(N-k)H\left(\frac{d}{N-k}\right)}$, and the largest term in
(\ref{eq:PEm1}) and (\ref{eq:PEm2}) is for
$d=\lfloor(N-k)(\mu_m-\epsilon_1)\rfloor$ or
$d=\lceil(N-k)(\mu_m+\epsilon_1)\rceil$, we have
\begin{eqnarray}
\textrm{Pr}\left(\mathbb{E}_m^c\right)\leq(N-k+1)2^{-(N-k)
\lambda_m(\epsilon_1)},
\end{eqnarray}
where $\lambda_m(\epsilon_1)=\min\left\{D_{KL}
\left(\mathcal{P}_{\mu_m-\epsilon_1}||\mathcal{P}_{\mu_m}\right),
D_{KL}\left(\mathcal{P}_{\mu_m+\epsilon_1}||\mathcal{P}_{\mu_m}
\right)\right\}>0$ for $\epsilon_1>0$.

Moreover, $\mu_m$ with $0\leq \mu_m\leq 1$ for $m\notin\varphi_1$
is also i.i.d. with mean
\begin{eqnarray}
\textrm{E}[\mu_m]&&\!\!\!\!\!\!\!\!\!\!
=\textrm{E}\left[\prod_{j\in\varphi^*_2}(1-X_{m,j}q)\right]\\
&&\!\!\!\!\!\!\!\!\!\!=\prod_{j\in\varphi^*_2}(1-\textrm{E}[X_{m,j}]q)\\
&&\!\!\!\!\!\!\!\!\!\!=(1-q^2)^{k^\prime}.
\end{eqnarray}
Let $\mathbb{E}$ be the event of
$\frac{1}{N-k}\sum_{m\notin\varphi_1} \mu_m\in
\mathcal{A}=\left\{d\left||d-(1-q^2)^{k^\prime}|\right.<\epsilon_2
\right\}$. Since the variance of $\mu_m$ is bounded by
$\textrm{V}[\mu_m]\leq 1$, using Chebyshev's inequality, we have
$\textrm{Pr}(\mathbb{E}^c)\leq \frac{1}{(N-k)\epsilon_2^2}$.

Let $\epsilon=\epsilon_1+\epsilon_2$. Let $\mathbb{E}^*$ be the
event of
$\{|\tilde{\alpha}_{sp}^N(\varphi)-1+(1-q^2)^{k^\prime}|<\epsilon\}$
when $|\varphi^*|=k^\prime$. To prove the lemma, we only need to
show $\lim_{N\rightarrow\infty}\textrm{Pr}(\mathbb{E}^*)=1$. The
law of total probability leads to
\begin{eqnarray}
\textrm{Pr}(\mathbb{E}^*)&&\!\!\!\!\!\!\!\!\!\!=\textrm{Pr}
\left(\cap_{m\notin\varphi_1}\mathbb{E}_m\cap\mathbb{E}\right)\textrm{Pr}
\left(\mathbb{E}^*|\cap_{m\notin\varphi_1}\mathbb{E}_m\cap\mathbb{E}\right)
\nonumber\\
&&\!\!\!\!\!\!\!\!\!\!\ \ \ +\textrm{Pr}
\left(\cup_{m\notin\varphi_1}\mathbb{E}_m^c\cup\mathbb{E}^c\right)
\textrm{Pr}\left(\mathbb{E}^*|\cup_{m\notin\varphi_1}\mathbb{E}_m^c
\cup\mathbb{E}^c\right).\nonumber
\end{eqnarray}
Using union bound
\begin{eqnarray}
&&\!\!\!\!\!\!\!\!\!\!\lim_{N\rightarrow\infty}\textrm{Pr}
\left(\cup_{m\notin\varphi_1}\mathbb{E}_m^c\cup\mathbb{E}^c\right)\nonumber\\
\leq&&\!\!\!\!\!\!\!\!\!\!\lim_{N\rightarrow\infty}
\left(\sum_{m\notin\varphi_1}\textrm{Pr}\left(\mathbb{E}_m^c\right)+\textrm{Pr}\left(\mathbb{E}^c\right)\right)\nonumber\\
\leq&&\!\!\!\!\!\!\!\!\!\! \lim_{N\rightarrow\infty} \left(
(N-k+1)^22^{-(N-k)\min_m\lambda_m(\epsilon_1)}
+ \frac{1}{(N-k)\epsilon_2^2}\right)\nonumber\\
=&&\!\!\!\!\!\!\!\!\!\!0.\nonumber
\end{eqnarray}
Thus, $\lim_{N\rightarrow\infty}\textrm{Pr}
\left(\cap_{m\notin\varphi_1}\mathbb{E}_m\cap\mathbb{E}\right)=1$.
Since $\mathbb{E}_m, m\notin\varphi_1$, and $\mathbb{E}$ directly
lead to $\mathbb{E}^*$, we have
$\textrm{Pr}\left(\mathbb{E}^*|\cap_{m\notin\varphi_1}\mathbb{E}_m
\cap\mathbb{E}\right)=1$. Therefore,
$\lim_{N\rightarrow\infty}\textrm{Pr}(\mathbb{E}^*)=1$, which
proves the lemma. \myQED

\emph{Proof of Theorem~\ref{thm:main1}:} According to
Corollary~\ref{cor:Iq}, we can derive $P_{\mathcal{I}_q^\infty}$
based on $\widetilde{\mathcal{I}}_q^N$ instead of
$\mathcal{I}_q^N$.  Moreover,  according to
Lemma~\ref{lem:scattered}, the SF pattern is scattered with
probability 1. We first derive $P_{\mathcal{I}_q^\infty}$ for
given $\Phi=\varphi\in\overline{\Omega}_N^k$.  The main idea  is
that with the knowledge of SF pattern $\Phi$ and SF rows and
columns $X_{sf}$, the entries of readback signal array $Y^{N\times
N}$ become independent with mixed Gaussian distribution.

Define index sets
\begin{eqnarray}
\mathcal{S}_v&&\!\!\!\!\!\!\!\!\!\!
=\left\{(m,n)\left|\bigcup_{(i,j)\in\varphi}\right.X_{m,j}X_{i,j}X_{i,n}
=v, 0\leq m, n\leq N\right\}, v=0,1,\nonumber \\
\widetilde{\mathcal{S}}_v&&\!\!\!\!\!\!\!\!\!\!=\left\{(m,n)
\left|(m,n)\in\mathcal{S}_v, \right.m\notin
\varphi_1, n\notin\varphi_2\right\}, v=0,1,\nonumber\\
\mathcal{S}^{sp}_v&&\!\!\!\!\!\!\!\!\!\!=\left\{(m,n)
\left|(m,n)\in\mathcal{S}_v, m\in \varphi_1\right. \textrm{or}\
n\in\varphi_2\right\}, v=0,1.\nonumber
\end{eqnarray}
For simplicity, we define the following auxiliary random variables
 \begin{eqnarray}
 X^*&&\!\!\!\!\!\!\!\!\!\!\sim\textrm{Bernoulli} (q),\ Z^*\sim
 \mathcal{N}(0, \sigma^2),\\
Y^*_v&&\!\!\!\!\!\!\!\!\!\!= R_{ X^*}(v)+Z^*, v=0,1,
 \end{eqnarray}
and the related probability density distributions are $P_{Y^*_v|
X^*}(y)=f(y,R_{X^*}(v),\sigma)$ and
$P_{Y^*_v}(y)=\sum_{x=0,1}\lambda_xf(y,R_x(v),\sigma), v=0,1$.

Given $X^{N\times N}$ and $\Phi=\varphi$,  entries of $Y^{N\times
N}$ in (\ref{eq:Y}) are independent Gaussian variables with
$P_{Y_{m,n}|X^{N\times N},\Phi}(Y_{m,n}|X^{N\times
N},\varphi)=f(Y_{m,n}, R_{X_{m,n}}(v),\sigma)=f(Z_{m,n},
0,\sigma)$ for $(m,n)\in \mathcal{S}_v, v=0, 1$.
 Therefore,
\begin{eqnarray}
&&\!\!\!\!\!\!\!\!\!\!\lim_{N\rightarrow\infty}\frac{1}{N^2} \log
P_{Y^{N\times N}|X^{N\times N},\Phi}\left(Y^{N\times N}|X^{N\times
N},
\varphi\right)\nonumber\\
=&&\!\!\!\!\!\!\!\!\!\!-\lim_{N\rightarrow\infty}\frac{1}{N^2}
\sum_{v=0,1}\sum_{(m,n)\in\mathcal{S}_v}\log
\frac{1}{f(Z_{m,n},0,\sigma)}
\nonumber\\
\overset{p}{=}&&\!\!\!\!\!\!\!\!\!\!-\lim_{N\rightarrow\infty}
\sum_{v=0,1}\frac{|\mathcal{S}_v|}{N^2}\textrm{E}
\left[\log \frac{1}{f(Z^*,0,\sigma)}\right]\nonumber\\
=&&\!\!\!\!\!\!\!\!\!\!-\log\sqrt{2\pi e\sigma^2},\label{eq:eHz}
\end{eqnarray}
where (\ref{eq:eHz}) is due to $\textrm{E} \left[\log
(1/f(Z^*,0,\sigma))\right]=\log\sqrt{2\pi e\sigma^2}$ and
$\sum_{v=0,1}{|\mathcal{S}_v|}/{N^2}=1$.

Given $X_{sf},\Phi$, entries of $Y^{N\times N}$ are independent
random variables but with four different distributions.
Specifically, $Y_{m,n}$ with $(m,n)\in\widetilde{\mathcal{S}}_v$
are i.i.d. mixed Gaussian with probability density function
$\sum_{x=0,1}\lambda_xf(Y_{m,n}, R_x(v),\sigma)$ for $v=0,1$, and
$Y_{m,n}$ with $(m,n)\in\mathcal{S}^{sp}_v$ are independent
Gaussian with probability density function  $f(Y_{m,n},
R_{X_{m,n}}(v),\sigma)=f(Z_{m,n}, 0,\sigma)$ for $v=0, 1$.
Therefore,
\begin{eqnarray}
&&\!\!\!\!\!\!\!\!\!\!\lim_{N\rightarrow\infty}\frac{1}{N^2} \log
\frac{1}{P_{Y^{N\times N}|X_{sf},\Phi}\left(Y^{N\times N}|X_{sf},
\varphi\right)}\nonumber\\
=&&\!\!\!\!\!\!\!\!\!\!\lim_{N\rightarrow\infty}
\sum_{v=0,1}\frac{1}{N^2}\left(\sum_{(m,n)\in\widetilde{\mathcal{S}}_v}
\log \frac{1}{\sum_{x=0,1}f(Y_{m,n}, R_x(v),\sigma)}\right.\nonumber\\
&&\!\!\!\!\!\!\!\!\!\!\ \ \ \ \ \ \ \ \  \ \ \ \ \ \ \ \  \left.
+\sum_{(m,n)\in\mathcal{S}^{sp}_v}\log \frac{1}{f(Z_{m,n},
0,\sigma)}\right)
\nonumber\\
\overset{p}{=}&&\!\!\!\!\!\!\!\!\!\!\lim_{N\rightarrow\infty}
\sum_{v=0,1}\left(\frac{|\widetilde{\mathcal{S}}_v|}{N^2}\textrm{E}
\left[\log \frac{1}{\sum_{x=0,1}
\lambda_xf(Y^*_v, R_x(v),\sigma)}\right]\right.\nonumber\\
&&\!\!\!\!\!\!\!\!\!\!\ \ \ \ \ \ \ \ \  \ \ \ \ \ \
+\left.\frac{|\mathcal{S}^{sp}_v|}{N^2}\textrm{E}
\left[\log \frac{1}{f(Z^*, 0,\sigma)}\right]\right)\label{eq:E4}\\
=&&\!\!\!\!\!\!\!\!\!\!\sum_{v=0,1}\lim_{N\rightarrow\infty}
\frac{|\widetilde{\mathcal{S}}_v|}{N^2}\textrm{E} \left[\log
\frac{1}{\sum_{x=0,1}\lambda_xf(Y^*_v, R_x(v), \sigma)}\right],
\label{eq:E2}
\end{eqnarray}
where (\ref{eq:E4}) is due to the law of large numbers and
(\ref{eq:E2}) is due to $|\mathcal{S}^{sp}_v|/N^2\leq
(2Nk-k^2)/N^2\rightarrow 0$. Using Lemma~\ref{lem:SPrate},
\begin{eqnarray}
\lim_{N\rightarrow\infty}\frac{|\widetilde{\mathcal{S}}_1|}{N^2}\
=&&\!\!\!\!\!\!\!\!\!\!\lim_{N\rightarrow\infty}
\frac{|\mathcal{S}_1|}{N^2}
=\lim_{N\rightarrow\infty}\alpha^N_{sp}(\varphi)\overset{p}{=}
\alpha_k,\label{eq:alpha}\\
\lim_{N\rightarrow\infty}\frac{|\widetilde{\mathcal{S}}_0|}{N^2}\overset{p}{=}
&&\!\!\!\!\!\!\!\!\!\!1-\alpha_k.\label{eq:1-alpha}
\end{eqnarray}
Since $\textrm{E}\left[\log{1/\sum_{x=0,1}\lambda_xf(Y^*_v,
R_x(v),\sigma)} \right]-\log\sqrt{2\pi e\sigma^2}=C_q(\gamma)$ and
$C_q(\gamma^\prime)$ for $v=0$ and $1$, applying (\ref{eq:alpha}),
(\ref{eq:1-alpha}) to (\ref{eq:E2}) and combing with
(\ref{eq:eHz}), we have
\begin{equation}
\lim_{N\rightarrow\infty}\widetilde{\mathcal{I}}_q^N
\overset{p}{=}C_q(\gamma^\prime)+(1-\alpha_k)(C_q(\gamma)
-C_q(\gamma^\prime)).
\end{equation}
Using the probability distribution function $P_{\alpha_k}$ derived
in Lemma~\ref{lem:SPrate}, we have
 \begin{eqnarray}
P_{\mathcal{I}_q^\infty}(x)=&&\!\!\!\!\!\!\!\!\!\!
\sum_{k^\prime=0}^k\binom{k}{k^\prime}q^{k^\prime}(1-q)^{k-k^\prime}\delta
\left[x-C_q(\gamma^\prime)\right.\nonumber\\
&& \ \ \ \  \left.-(1-q^2)^{k^\prime}\left(C_q(\gamma)
-C_q(\gamma^\prime)\right)\right]
\end{eqnarray}
for given $\Phi=\varphi\in\overline{\Omega}_N^k$. Weighted by the
probability of $\overline{\Omega}^k, k=0,1,...,K$, derived by
Lemma~\ref{lem:scattered}, we obtain  (\ref{eq:PI}).
Theorem~\ref{thm:main1} is proved. \myQED

The proof of Theorem~\ref{thm:main1} also inspires us to conceive
new decoding schemes for ReRAM. Since if the SF pattern $\Phi$ and
the SF rows and columns of $X_{sf}$ are known, the entries of
$Y^{N\times N}$ become uncorrelated, we can first decode $X_{sf}$
based on $Y^{N\times N}$, which is easy when the array is large,
and then decode the remaining data of $X_{sf}^c$ as a memoryless
channel. The initial attempt of this idea was presented in our
recent work \cite{SongArxiv}.

\section{Concluding Remarks}\label{sec:conclude}
In this paper, we proposed a maximum achievable rate  analysis for
ReRAM channel with i.i.d. Bernoulli $(q)$ inputs. The main
challenge is the channel correlation caused by sneak path
interference. We resort to the mutual information spectrum method
and derives the maximum achievable rate as an explicit function of
the probability distribution of the SF number and input
distribution $q$. ReRAM channels with different settings, such as
single- and across-array codings, optimal and TIN decodings, with
and without data shaping are compared. These results provide us
valuable insights on code design for ReRAM systems.

Since in our analysis the input distribution is subject to i.i.d.
Bernoulli $(q)$, the maximum achievable rate derived in this paper
may not be the capacity of ReRAM channel. To prove that it is
exactly the channel capacity, we still need to show that
(\ref{eq:R}) can not be exceeded by non-i.i.d. data distributions.
A common coding technique to generate such data input is the
constrained coding \cite{ChiLett,Xingwei,Immink}.

We used a bounded distribution of the SF number with maximum $K$
SFs in each array, rather than assuming that the SF occurs
independently with a fixed probability for each cell (as assumed
in \cite{Ben,CZH,SongTcom}). It should be mentioned that the
assumption in this work is more meaningful for both of the maximum
achievable rate analysis and practical code design. The reason is
that the assumption in \cite{Ben,CZH,SongTcom} leads to a linear
increase of SF number with the array dimension $N^2$, which will
approach infinity as $N$ approaches infinity. Under this
assumption, the maximum achievable rate will always be
$C_{0.5}(\gamma^\prime)$. On the other hand, in practice, we need
a parameter such as the maximum tolerable SF number, (the worst
channel status) to guide the code design.

\appendices

\section{Proof of Lemma~\ref{lem:replace} (Second Equation)}
\label{lem1proof2}

Define the following sets for $x^{N\times N}\in \{0, 1\}^{N\times
N}, \varphi\in \Omega_N, y^{N\times N}\in \mathbb{R}^{N\times N}$,
\begin{eqnarray}
&&\!\!\!\!\!\!\!\!\!\! A_N^\prime=\nonumber\\
&&\!\!\!\!\!\!\!\!\!\!\left\{\left(x^{N\times N},\varphi,
y^{N\times N}\right)\left|\log\frac{P_{Y^{N\times N}|X^{N\times
N}} \left(y^{N\times N}|x^{N\times N}\right)}{P_{Y^{N\times
N}|x^{N\times N}, \Phi}\left(y^{N\times N}|x^{N\times
N},\varphi\right)}
\leq\frac{-\epsilon N^2}{2}\right.\right\}\nonumber\\
&&\!\!\!\!\!\!\!\!\!\! B_N^\prime=\nonumber\\
&&\!\!\!\!\!\!\!\!\!\!\left\{\left(x^{N\times N},\varphi,
y^{N\times N}\right)\left|\log\frac{P_{Y^{N\times N}|X^{N\times
N}} \left(y^{N\times N}|x^{N\times N}\right)}{P_{Y^{N\times
N}|X^{N\times N}, \Phi}\left(y^{N\times N}|x^{N\times
N},\varphi\right)} \geq\frac{\epsilon
N^2}{2}\right.\right\}\nonumber.
\end{eqnarray}
Since
\begin{eqnarray}
&&\!\!\!\!\!\!\!\!\!\!\textrm{Pr}\left(\frac{1}{N^2}
\left|\log\frac{P_{Y^{N\times N}|X^{N\times N}} \left(Y^{N\times
N}|X^{N\times N}\right)}{P_{Y^{N\times N}|X^{N\times N},
\Phi}\left(Y^{N\times N}|X^{N\times N},\Phi\right)}\right|
<\frac{\epsilon}{2}\right)\nonumber\\ =&&\!\!\!\!\!\!\!\!\!\!1
-\sum_{(x^{N\times N},\varphi, y^{N\times N})\in A_N^\prime \cup
B_N^\prime}P_{X^{N\times N},\Phi,Y^{N\times N}} \left(x^{N\times
N},\varphi, y^{N\times N}\right),\nonumber
\end{eqnarray}
to prove the second equation of Lemma~\ref{lem:replace}, we only
need to show that
\begin{equation}
\sum_{(x^{N\times N},\varphi, y^{N\times N})\in A_N^\prime \cup
B_N^\prime}P_{x^{N\times N},\Phi,Y^{N\times N}} \left(x^{N\times
N},\varphi,y^{N\times N}\right)\rightarrow0.\nonumber
\end{equation}

Using the fact that the joint probability is smaller than or equal
to the marginal probability, we have
\begin{eqnarray}
&&\!\!\!\!\!\!\!\!\!\!\sum_{(x^{N\times N},\varphi, y^{N\times N})
\in A_N^\prime}P_{X^{N\times N},\Phi,Y^{N\times
N}}\left(x^{N\times N},
\varphi,y^{N\times N}\right)\nonumber\\
\leq&&\!\!\!\!\!\!\!\!\!\! \sum_{(x^{N\times N},\varphi,
y^{N\times N}) \in A_N^\prime}P_{X^{N\times N},Y^{N\times
N}}\left(x^{N\times N},
y^{N\times N}\right)\nonumber\\
\leq&&\!\!\!\!\!\!\!\!\!\! \sum_{(x^{N\times N},\varphi,
y^{N\times N}) \in A_N^\prime}P_{X^{N\times N}}\left(x^{N\times
N}\right) P_{Y^{N\times N}|X^{N\times N}}\left(y^{N\times
N}|x^{N\times N}\right)
\nonumber\\
\leq&&\!\!\!\!\!\!\!\!\!\! e^{-\epsilon N^2/2}\sum_{(x^{N\times
N}, \varphi, y^{N\times N})\in A_N^\prime}P_{X^{N\times
N}}\left(x^{N\times N} \right)P_{Y^{N\times N}|X^{N\times N},\Phi}
\left(y^{N\times N}|x^{N\times N},\varphi\right)\nonumber\\
\leq&&\!\!\!\!\!\!\!\!\!\! e^{-\epsilon N^2/2}\sum_{\varphi\in
\Omega_N}\sum_{x^{N\times N}\in \{0, 1\}^{N\times N}}P_{X^{N\times
N}}
\left(x^{N\times N}\right)\nonumber\\
&&\!\!\!\!\!\!\!\!\!\!\ \ \ \ \ \ \ \ \ \ \ \  \ \ \ \  \times
\sum_{y^{N\times N}\in\mathbb{R}^{N\times N}} P_{Y^{N\times
N}|X^{N\times N},\Phi}\left(y^{N\times N}|x^{N\times N},
\varphi\right)\nonumber\\
<&&\!\!\!\!\!\!\!\!\!\!e^{-\epsilon N^2/2} (K+1)N^{2K}\rightarrow
0, \nonumber
\end{eqnarray}
where we used $|\Omega_N|\leq (K+1)N^{2K}$, $\sum_{x^{N\times
N}\in \{0, 1\}^{N\times N}}P_{X^{N\times N}}\left(x^{N\times
N}\right)=1$, and the fact that for any pair of $(x^{N\times
N},\varphi)$, $\sum_{y^{N\times N}\in \in\mathbb{R}^{N\times
N}}P_{Y^{N\times N} |X^{N\times N},\Phi}\left(y^{N\times
N}|x^{N\times N},\varphi\right)=1$.

Moreover,
\begin{eqnarray}
&&\!\!\!\!\!\!\!\!\!\!\sum_{(x^{N\times N},\varphi, y^{N\times N})
\in B_N^\prime}P_{X^{N\times N},\Phi,Y^{N\times
N}}\left(x^{N\times N},
\varphi,y^{N\times N}\right)\nonumber\\
=&&\!\!\!\!\!\!\!\!\!\! \sum_{(x^{N\times N},\varphi, y^{N\times
N}) \in B_N^\prime}P_{X^{N\times N},\Phi}\left(x^{N\times
N},\varphi\right) P_{Y^{N\times N}|X^{N\times
N},\Phi}\left(y^{N\times N}|x^{N\times N},
\varphi\right)\nonumber\\
\leq&&\!\!\!\!\!\!\!\!\!\! e^{-\epsilon N^2/2}\sum_{(x^{N\times
N}, \varphi, y^{N\times N})\in B_N^\prime}P_{X^{N\times N},\Phi}
\left(x^{N\times N},\varphi\right)P_{Y^{N\times N}|X^{N\times N}}
\left(y^{N\times N}|x^{N\times N}\right)\nonumber\\
\leq&&\!\!\!\!\!\!\!\!\!\! e^{-\epsilon N^2/2}\sum_{x^{N\times N}
\in\{0 ,1\}^{N\times N},\varphi\in\Omega_N}P_{X^{N\times N},\Phi}
\left(x^{N\times N},\varphi\right)\nonumber\\
&&\!\!\!\!\!\!\!\!\!\!\ \ \ \ \ \ \ \ \ \ \ \ \   \ \ \ \times
\sum_{y^{N\times N}\in\mathbb{R}^{N\times N}} P_{Y^{N\times
N}|X^{N\times N}}\left(y^{N\times N}|x^{N\times N}\right)
\nonumber\\
=&&\!\!\!\!\!\!\!\!\!\! e^{-\epsilon N^2/2}\rightarrow 0\nonumber
\end{eqnarray}
where we used $\sum_{x^{N\times N}\in\{0 ,1\}^{N\times N},
\varphi\in\Omega_N}P_{X^{N\times N},\Phi}\left(x^{N\times N},
\varphi\right)=1$, and the fact that for any $x^{N\times N}$, it
holds that $\sum_{y^{N\times N}\in \in\mathbb{R}^{N\times N}}
P_{Y^{N\times N}|X^{N\times N}}\left(y^{N\times N}|x^{N\times
N}\right)=1$. The second equation of Lemma~\ref{lem:replace} is
proved.

\begin{IEEEbiographynophoto}{Guanghui Song}
    (M13) received the B.E. degree in communication and engineering
    from Henan Normal University, Xinxiang, China, in 2006. He received
    the M.S. degree in telecommunications engineering from Xidian
    University, Xi'an, China, in 2009 and the Ph.D. degree in the
    department of intelligent information engineering and sciences,
    Doshisha University, Kyoto, Japan, in 2012. From 2013 to 2021,
    he did postdoctoral research in Doshisha University, Kyoto, Japan,
    University of Western Ontario, London, Canada, and Singapore
    University of Technology and Design, Singapore. Currently, he
    is an Associate Professor with Xidian University, Xi'an, China.
    His research interests are in the areas of channel coding,
    multi-user coding, and coding for data storage systems.
\end{IEEEbiographynophoto}

\begin{IEEEbiographynophoto}{Kui Cai}
    (SM11)  received the B.E. degree in information and
    control engineering from Shanghai Jiao Tong University,
    Shanghai, China, and joint Ph.D. degree in electrical
    engineering from Technical University of Eindhoven, The
    Netherlands, and National University of Singapore. Currently,
    she is an Associate Professor with Singapore University of
    Technology and Design (SUTD). She received 2008 IEEE
    Communications Society Best Paper Award in Coding and
    Signal Processing for Data Storage. She is an IEEE senior
    member, and served as the Vice-Chair (Academia) of IEEE
    Communications Society, Data Storage Technical Committee
    (DSTC) during 2015 and 2016. Her main research interests
    are in the areas of coding theory, information theory, and
    signal processing for various data storage systems and digital
    communications.
\end{IEEEbiographynophoto}

\begin{IEEEbiographynophoto}{Ying Li}(M08)
    received the B.E. degree
    in telecommunication engineering and the Ph.D.
    degree in communication and information systems
    from Xidian University, Xian, China, in 1995 and
    2005, respectively. From 2011 to 2012, she was
    with the University of California at Davis, Davis,
    CA, USA, as a Visiting Scholar. She is currently
    a Professor with Xidian University. Her current
    research interests include design and analysis for
    wireless system, including channel coding, wireless
    network communications, interference processing,
    and MIMO techniques.
\end{IEEEbiographynophoto}

\begin{IEEEbiographynophoto}{Kees A. Schouhamer Immink} (M81-SM86-F90)
    founded
    Turing Machines Inc. in 1998, an innovative start-up focused
    on novel signal processing for DNA-based storage, where he
    currently holds the position of president. He was from 1994
    till 2014 an adjunct professor at the Institute for Experimental
    Mathematics, Essen-Duisburg University, Germany.

    He contributed to digital video, audio, and data recording
    products including Compact Disc, CD-ROM, DCC, DVD,
    and Blu-ray Disc. He received the 2017 IEEE Medal of
    Honor, a Knighthood in 2000, a personal Emmy award in
    2004, the 1999 AES Gold Medal, the 2004 SMPTE Progress
    Medal, the 2014 Eduard Rhein Prize for Technology, and the
    2015 IET Faraday Medal. He received the Golden Jubilee
    Award for Technological Innovation by the IEEE Information
    Theory Society in 1998. He was inducted into the Consumer
    Electronics Hall of Fame, elected into the Royal Netherlands
    Academy of Sciences and the (US) National Academy of
    Engineering. He received an honorary doctorate from the
    University of Johannesburg in 2014. He served the profession
    as President of the Audio Engineering Society inc., New
    York, in 200
\end{IEEEbiographynophoto}

\end{document}